\documentclass[prd,aps,floatfix,preprint,nofootinbib,11pt]{article}
\usepackage{graphicx,color}
\usepackage{epstopdf}
\usepackage{amssymb}
\usepackage{amsmath}
\usepackage{subfigure}
\usepackage{epsfig}
\usepackage{float}
\usepackage{bm}
\usepackage{jheppub} 
\usepackage[T1]{fontenc} 
\usepackage{array}
\usepackage{hyperref}

\def\simge{\mathrel{%
       \rlap{\raise 0.511ex \hbox{$>$}}{\lower 0.511ex \hbox{$\sim$}}}}
\def\simle{\mathrel{
       \rlap{\raise 0.511ex \hbox{$<$}}{\lower 0.511ex \hbox{$\sim$}}}}

\makeatletter

\newcommand*{\doi}[1]{\href{https://doi.org/#1}{doi:\ #1}}
\newcommand{\figcaption}[1]{\def\@captype{figure}\caption{#1}}
\newcommand{\tblcaption}[1]{\def\@captype{table}\caption{#1}}

\newcommand{\no}{\nonumber}

\newcommand{\hw}{\hat{w}}

\newcommand{\hsg}{\hat{s}_g}
\newcommand{\ho}{\hat{o}}
\newcommand{\homg}{\hat{\omega}}
\newcommand{\hQ}{\hat{Q}}

\newcommand{\DU}{{\cal D}U}

\newcommand{\Nsite}{N_{\rm site}}

\newcommand{\la}{\left\langle\,}
\newcommand{\ra}{\,\right\rangle}

\renewcommand{\th}{\theta}

\newcommand{\x}{\vec x}
\newcommand{\dis}{\displaystyle}
\newcommand{\tR}{\theta_R}
\newcommand{\tI}{\theta_I}
\newcommand{\eps}{\varepsilon}

\makeatother

\title{
$\th$ dependence of $T_c$ in 4d SU(3) Yang-Mills theory with histogram
method and the Lee-Yang zeros in the large $N$ limit
}

\author[a]{Noriaki~Otake}
\emailAdd{noriotak@post.kek.jp}
\author[a,b]{and\ Norikazu~Yamada}
\emailAdd{norikazu.yamada@kek.jp}
\affiliation[a]{
        Graduate University for Advanced Studies (SOKENDAI), %
        Tsukuba 305-0801, Japan}
\affiliation[b]{
        High Energy Accelerator Research Organization (KEK), %
        Tsukuba 305-0801, Japan}
\date{\today}

\abstract{
 The phase diagram on the $\th$-$T$ plane in four dimensional SU(3)
 Yang-Mills theory is explored.
 We revisit the $\th$ dependence of the deconfinement transition
 temperature, $T_c(\th)$, on the lattice through the  constraint
 effective potential for Polyakov loop.
 The $\th$ term is introduced by the reweighting method, and the
 critical $\beta$ is determined to $\th \sim 0.75$, where the
 interpolation in $\beta$ is carried out by the multipoint reweighting
 method.
 The $\th$ dependence of $T_c$ obtained here turns out to be consistent
 with the previous result by D'Elia and
 Negro~\cite{DElia:2012pvq,DElia:2013uaf}.
 We also derive $T_c(\th)$ by classifying configurations into the high
 and low temperature phases and applying the Clausius-Clapeyron
 equation.
 It is found that the potential barrier in the double well potential at
 $T_c(\th)$ becomes higher with $\th$, which suggests that the first
 order transition continues robustly above $\th \sim 0.75$.
 Using information obtained here, we try to depict the expected $\th$
 dependence of the free energy density at $T\simle T_c(0)$, which
 crosses the first order transition line at an intermediate value of
 $\th$.
 Finally, how the Lee-Yang zeros associated with the spontaneous CP
 violation appear is discussed formally in the large $N$ limit, and the
 locations of them are found to be
 $(\tR,\tI)=\left( (2m+1)\pi, \frac{2n+1}{2\chi V_4} \right)$ with $m$
 and $n$ arbitrary integers.
}

\begin{document}
\maketitle

\section{Introduction}
\label{sec:introduction}

To understand phases realized in a theory and how those change under
varying parameters is one of the most attractive aspects of field
theory.
In this work, we consider the $\th$-$T$ phase diagram for four
dimensional SU(3) Yang-Mills theory, where $\th$ is a parameter
controlling relative weights of different topological sectors in the path
integral~\cite{Callan:1976je} and $T$ denotes temperature.

In the high temperature deconfined phase, it has been known that
instanton
calculus~\cite{Polyakov:1975rs,Belavin:1975fg,Harrington:1978ve} is
reliable~\cite{Gross:1980br,Frison:2016vuc}, and no phase transition is
expected to occur by changing $\th$.
Interestingly, in the low temperature phase, it has been argued that
spontaneous CP violation takes place at $\th=\pi$ if the vacuum is in
the confined phase there~\cite{Gaiotto:2017yup,Kitano:2017jng}.
In the large $N$ limit, the occurrence of the CP violation at $\th=\pi$
seemed to be
established~\cite{tHooft:1973alw,Witten:1980sp,tHooft:1981bkw,Witten:1998uka}.
It is speculated based on numerical evidences that CP symmetry is also
broken at $\th=\pi$ in the opposite limit, {\it i.e.} SU(2) Yang-Mills
theory~\cite{Kitano:2020mfk}.
Recently, the $\th$ dependence of the vacuum energy density of SU(2)
theory was calculated on the lattice for $\th\simle 3\pi/2$, and the
spontaneous CP violation at $\th=\pi$ is concluded
in~\cite{Kitano:2021jho}, where a newly developed method, called the
subvolume method, is applied.

Although it is straightforward to apply the subvolume method to explore
the phase diagram of SU(3) Yang-Mills theory, it requires some cares.
Since the subvolume method can be seen as a variant of the reweighting
method, it may suffer from the so-called overlap problem.
In the SU(3) case, the deconfinement transition is of first order, and
when used to study phases with varying $\th$ the method may fail to
detect the phase transition and follow the original branch even after
passing the transition point~\footnote{Indeed,
the subvolume method could not detect the first order phase transition
associated with spontaneous CP violation at $\th=\pi$ and sticks to the
branch in the confined phase even after passing
$\th=\pi$~\cite{Kitano:2021jho}.}.

One approach complementary to calculating the free energy density with
the subvolume method is to identify the curve of $T_c(\th)$ in the phase
diagram.
Especially, if one could have succeeded to determine $T_c(\th)$ to
$\th=\pi$, it becomes clear whether $T_c(\th)$ touches to the $T=0$ axis
or not.
Unfortunately, it is difficult to estimate the critical temperature at
general values of $\th$ because the standard lattice techniques do not
work in the region with $\th \simge O(1)$, and hence as a first step we
constrain our discussion to the small $\th$ region, where several
numerical techniques are available.
In~\cite{DElia:2012pvq,DElia:2013uaf}, $T_c(\th)$ is determined in such
a region with the analytic continuation from imaginary $\th$ and the
reweighting method by monitoring the Polyakov loop susceptibility and
found to decrease with $\th$.
The aims of this work are to confirm this $\th$ dependence and obtain
different insights through a different approach.

In this work, we construct the constraint effective
potential~\cite{ORaifeartaigh:1986axd} from the histogram of the
Polyakov loop to identify $T_c(\th)$.
The $\th$ term is introduced by the standard reweighting technique.
With the effective potential at hand, we hope to gain some insights on
how the potential shape changes or whether the first order transition at
$T_c(\th)$ becomes stronger or weaker with $\th$.
Furthermore, according to the resulting histogram it becomes possible to
classify each configuration into the high and low temperature phases.
Using those configurations, we can separately calculate the free energy
density in two phases and combine them to depict the $\th$ dependence of
the free energy density across the first order phase transition.

In the discussion of the phase transition, the zeros of partition
function~\cite{Yang:1952be,Lee:1952ig,Fischer:1965rna} are often
analyzed.
On the $\th$-$T$ plane, at least, two kinds of phase transitions exist
corresponding to the center and CP symmetry breaking, respectively.
After briefly recalling how the zeros appear in the deconfinement
transition, we perform a formal discussion to explore the locations of
them associated with the spontaneous CP violation at $\th=\pi$ in the
large $N$ limit.

In sec.~\ref{sec:setup}, we briefly describe the lattice setup and the
methods as well as some basic results to show the features of the
ensembles used in the following analysis.
In sec.~\ref{sec:results}, the details of the numerical analysis
and the results for $T_c(\th)$ are presented, where the consistencies
with the previous result and the Clausius-Clapeyron equation are tested.
The analyses using the configurations separated into high and low
temperature phases are also given here.
In sec.~\ref{sec:lee-yang-zeros}, the Lee-Yang zeros associated with
phase transitions relevant to the present study are explored.
Finally, the summary and the outlook are stated in
sec.~\ref{sec:summary}.
In Appendix \ref{app:cc}, the application of the Clapeyron-Clausius
equation to the determination of $dT(\th)/d\th\big|_{\th=0}$ is
described.

\section{Lattice Setup and method}
\label{sec:setup}
\subsection{parameters}

The partition function for lattice SU($N$) Yang-Mills theory including
the $\th$ term is
\begin{align}
   Z(\beta,\th)
=  \int\!\DU\, e^{-6\,\beta\,N_{\rm site}\,\hsg-i\th\hQ}\ ,\qquad
\end{align}
where $\hsg$ is the action density averaged over the number of the
lattice sites, $\Nsite$, and given by the sum of the average of the
plaquette $\hw_P$ and the rectangle $\hw_R$,
\begin{align}
   \hsg
=&\
   c_0(1-\hw_P)+2\,c_1(1-\hw_R)
\ ,
\end{align}
where $c_0$ and $c_1$ satisfying $c_0=1-8c_1$ are the improvement
coefficients for the lattice gauge action and $c_1=-0.331$ is taken
corresponding to the Iwasaki RG improved action~\cite{Iwasaki:1985we}.
The lattice gauge coupling is tuned by changing $\beta=2N/g^2$.
Throughout the paper, a hat $\ \hat{}\ $  is attached to functions of link
variable, {\it e.g.} $\hsg=\hsg[U]$, to distinguish from $c$-numbers
without a hat $\ \hat{}\ $.

In the definition of the topological charge on the lattice
$\hQ=\sum_x\hat{q}(x)$, we employ the five-loop improved topological
charge density operator $\hat{q}(x)$~\cite{deForcrand:1997esx}.
The topological charge for a given configuration is calculated after
$n_{\rm APE}$ times of the APE smearing~\cite{Albanese:1987ds}, and
observables are extrapolated to $n_{\rm APE}=0$ using those obtained in
the range of $n_{\rm APE}=[35, 55]$, which is chosen following the
criterion given in \cite{Kitano:2020mfk}.
Since the $n_{\rm APE}$ dependence for any observables studied here
turns out to be negligibly small, we will present results without
specifying $n_{\rm APE}$.

The number of lattice sites is $\Nsite=N_S^3\times N_T$ with $N_S=24$
and $N_T=6$.
We choose four values of $\beta$ with $N=3$ to cover the critical
$\beta$ at $\th=0$, $\beta_c\sim 2.515$~\cite{Okamoto:1999hi}.
The number of configurations at each $\beta$ is 40,000.
The statistical errors are estimated by the jackknife method.
Simulation parameters, $T/T_c$ and the statistics are summarized in
Tab.~\ref{tab:lat-sim}.
\begin{table}[tbp]
 \begin{center}
 \begin{tabular}{|c|cl|c|c|}
 \hline
& $\beta$ & $T/T_c$
& statistics 
\\
\hline
SU(3)
&  2.505 & 0.984 & 40,000 
\\
Iwasaki RG
&  2.510 & 0.992 & 40,000 
\\
$24^3\times 6$
&  2.515 & 1.000 & 40,000 
\\
&  2.520 & 1.008 & 40,000 
\\
\hline
 \end{tabular}
 \caption{
  The lattice parameters of the ensembles and other informations.
  }
 \label{tab:lat-sim}
 \end{center}
\end{table}

\subsection{constraint effective potential}

The histogram method or the constraint effective potential described
below is a useful tool especially when exploring phase boundaries and
has been used, for example, in the study of the phase diagram for many
flavor QCD \cite{Ejiri:2012rr} or in searching for the critical end
point in the heavy quark region~\cite{Ejiri:2019csa}.

The histogram for CP even operators $\ho_i$ ($i=1, 2, \cdots , N_o$)
measured at $\beta$ and $\th$ is defined by
\begin{align}
 p(o_1,\cdots,o_{N_o};\beta,\th)
&=\ \frac{1}{Z(\beta,\th)}\int\!\!\DU 
  \left\{\dis\prod_{i=1}^{N_o}\delta(\ho_{i}-o_{i})\right\}
  e^{-6\beta\Nsite s_g-i\th\hQ}
\no\\
&=\ { \la \left\{\dis\prod_{i=1}^{N_o}\delta(\ho_{i}-o_{i})\right\}
     \cos(\th\hQ)\ra_{\beta}
    \over
     \la \cos(\th\hQ)\ra_{\beta} }
\ ,
\label{eq:gene-hist}\\
   p(o_2,\cdots,o_{N_o};\beta,\th)
&=\ \int\!\! do_{1}\,p(o_1,\cdots,o_{N_o};\beta,\th)
\label{eq:int-hist}
\ ,
\end{align}
where $o_{i}$ denotes a c-number.
As indicated in \eqref{eq:gene-hist}, the $\th$ term is interpreted as a
part of observable, that is, introduced through the reweighting method.
$\la\cdots\ra_{\beta}$ denotes the expectation value over the
configurations generated at $\beta$ without the $\th$ term.

\begin{figure}[tb]
 \begin{center}
  \begin{tabular}{cc}
  \hspace*{-2ex}
  \includegraphics[width=0.5 \textwidth]
  {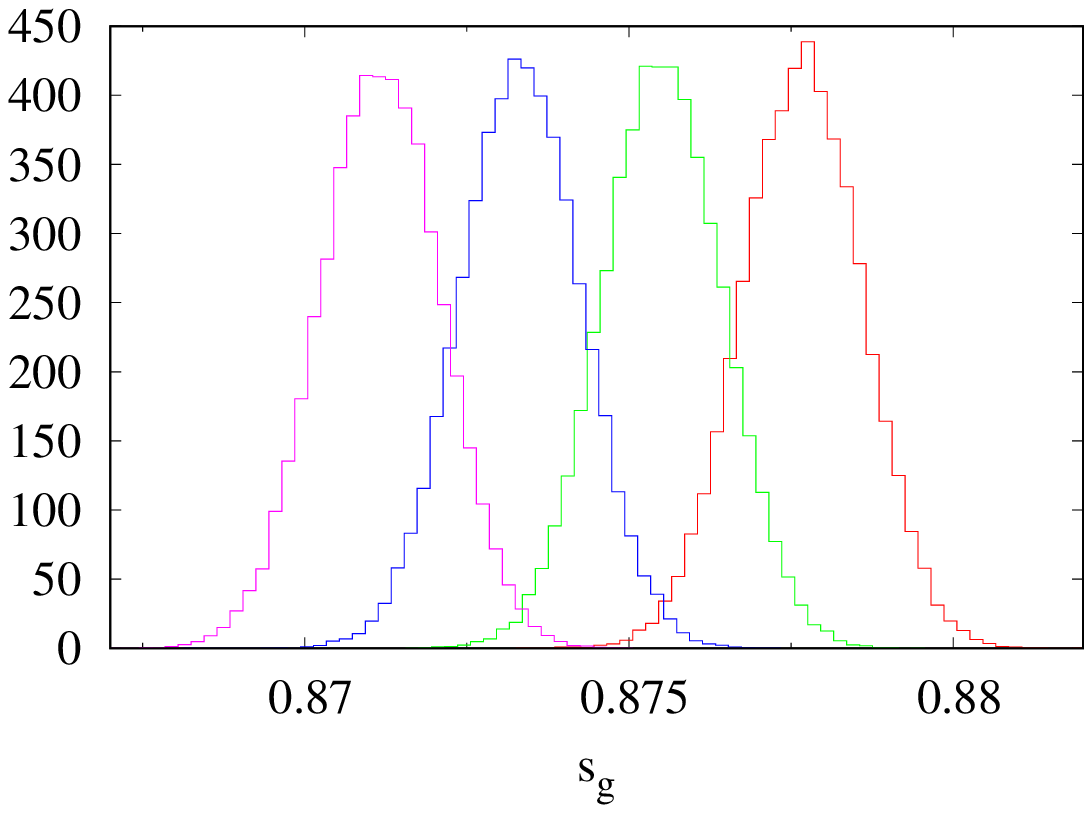} &
  \hspace{-2ex}
  \includegraphics[width=0.5 \textwidth]
  {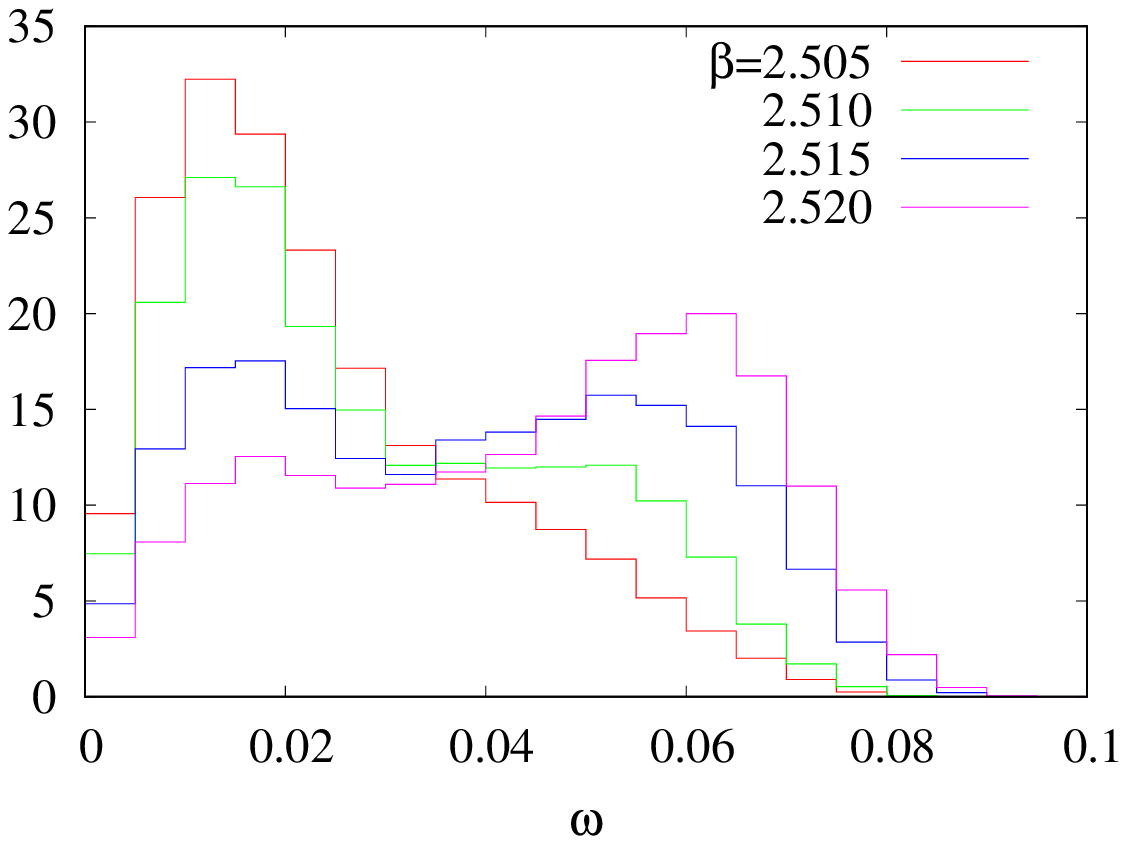}
\end{tabular}  
 \end{center}
 \caption{The histogram of the action $s_g$ (left) and the modulus of
 the Polyakov loop $\omega$ (right) for the four ensembles.
 }
 \label{fig:iws-hist-sg-omg}
\end{figure}

Some examples are shown in what follows because they would help to grab
the features of ensembles used in this work.
The first examples with $N_o=1$ shown in Fig.~\ref{fig:iws-hist-sg-omg}
are the histograms for the action density, $p(s_g;\beta,0)$, and the
modulus of the Polyakov loop, $p(\omega;\beta,0)$ at $\th=0$,
where
\begin{align}
 \homg
=\left|\hat{\Omega}\right|
=\left|\frac{1}{N_S^3\,N_c} \sum_{\x}
 {\rm Tr_c}\big[\dis\prod_{t=1}^{N_T} U_4(\x,t)\big]\right|\ ,
\end{align}
where $U_4(\x,t)$ is the link variable in time direction.
$\homg$ is an approximate order parameter for the
confinement-deconfinement transition.
In each plot, the four histograms correspond to the four values of
$\beta$.
Each histogram for $s_g$ well overlaps with, at least, one of others,
which becomes important later when interpolating histograms in $\beta$.
Some of the histograms for $\omega$ show a clear double-peak, which
indicates that the $\beta$ value for those are close to the critical
point.

\begin{figure}[tb]
 \begin{center}
  \begin{tabular}{cccc}
  \hspace*{-3ex}
  \includegraphics[width=0.25 \textwidth]
  {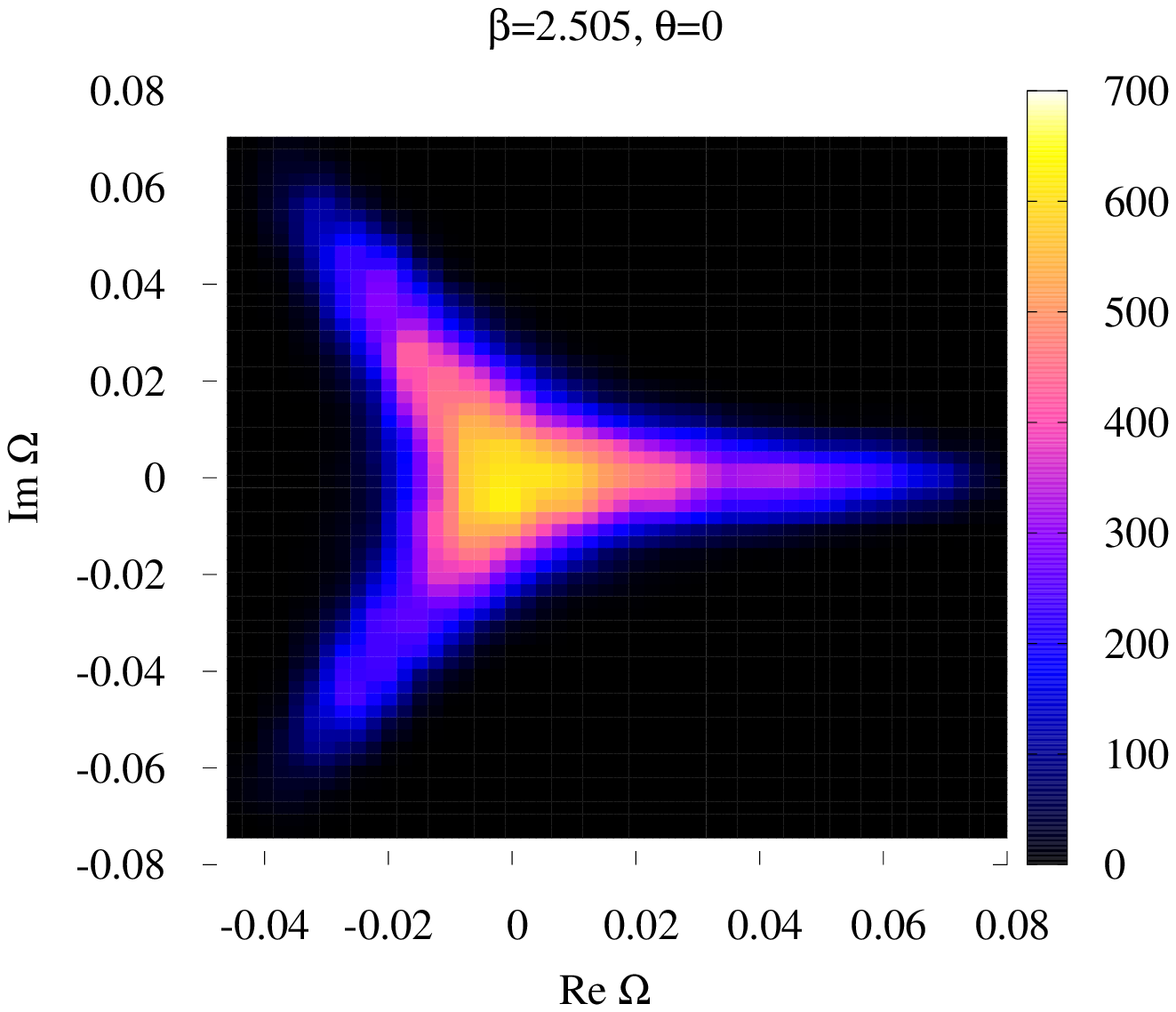} &
  \hspace{-3ex}
  \includegraphics[width=0.25 \textwidth]
  {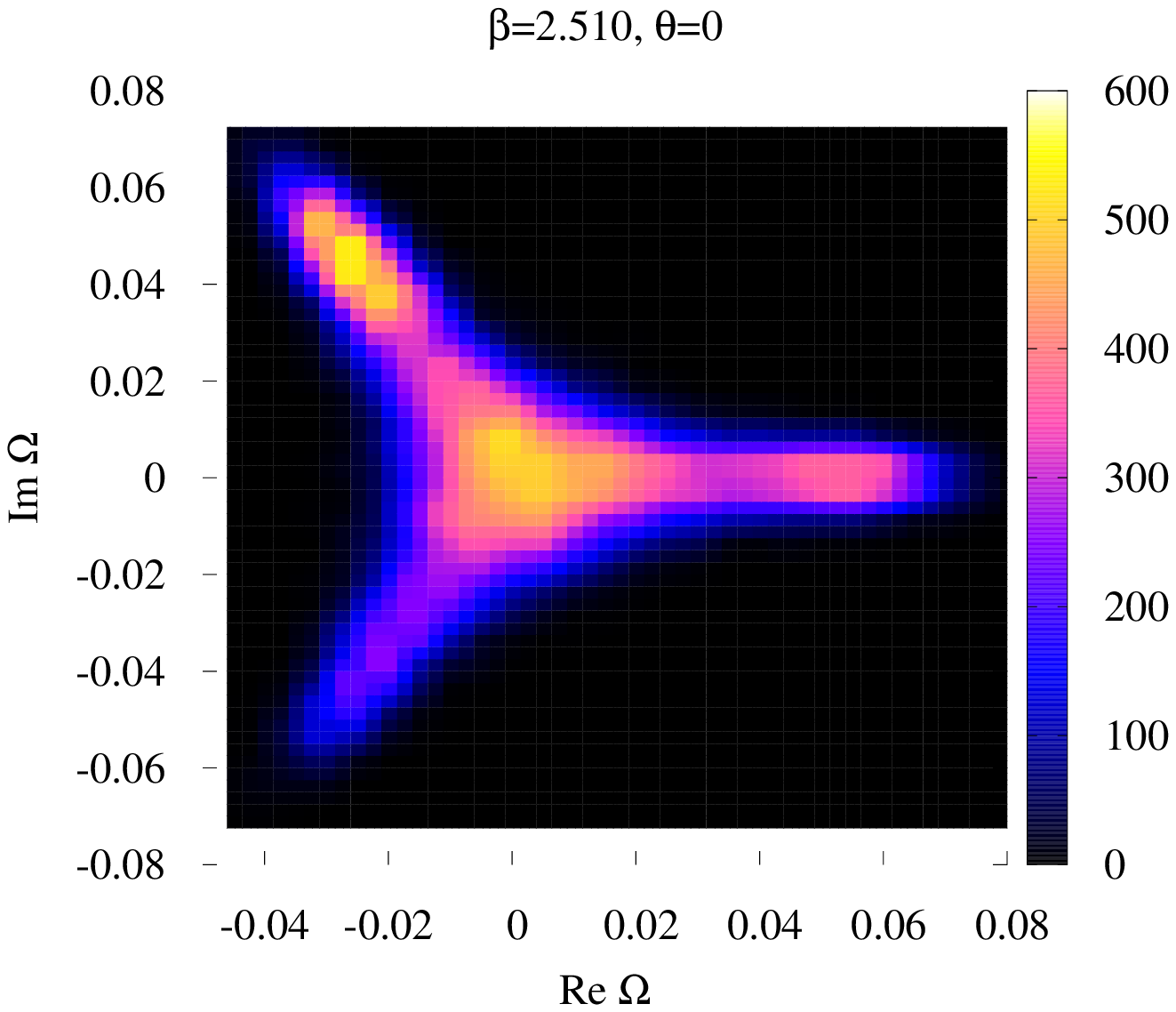} &
  \hspace{-3ex}
  \includegraphics[width=0.25 \textwidth]
  {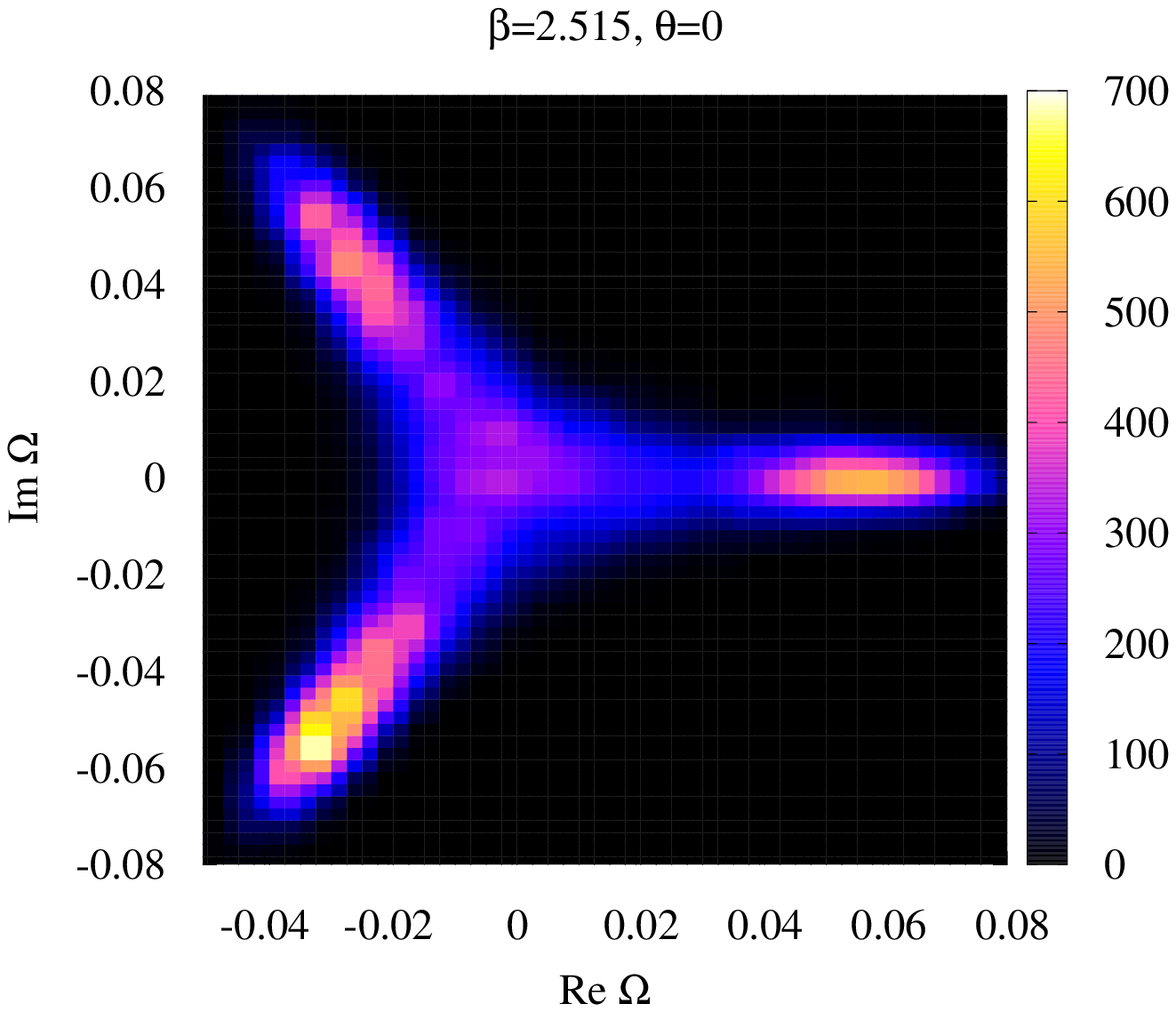} &
  \hspace{-3ex}
  \includegraphics[width=0.25 \textwidth]
  {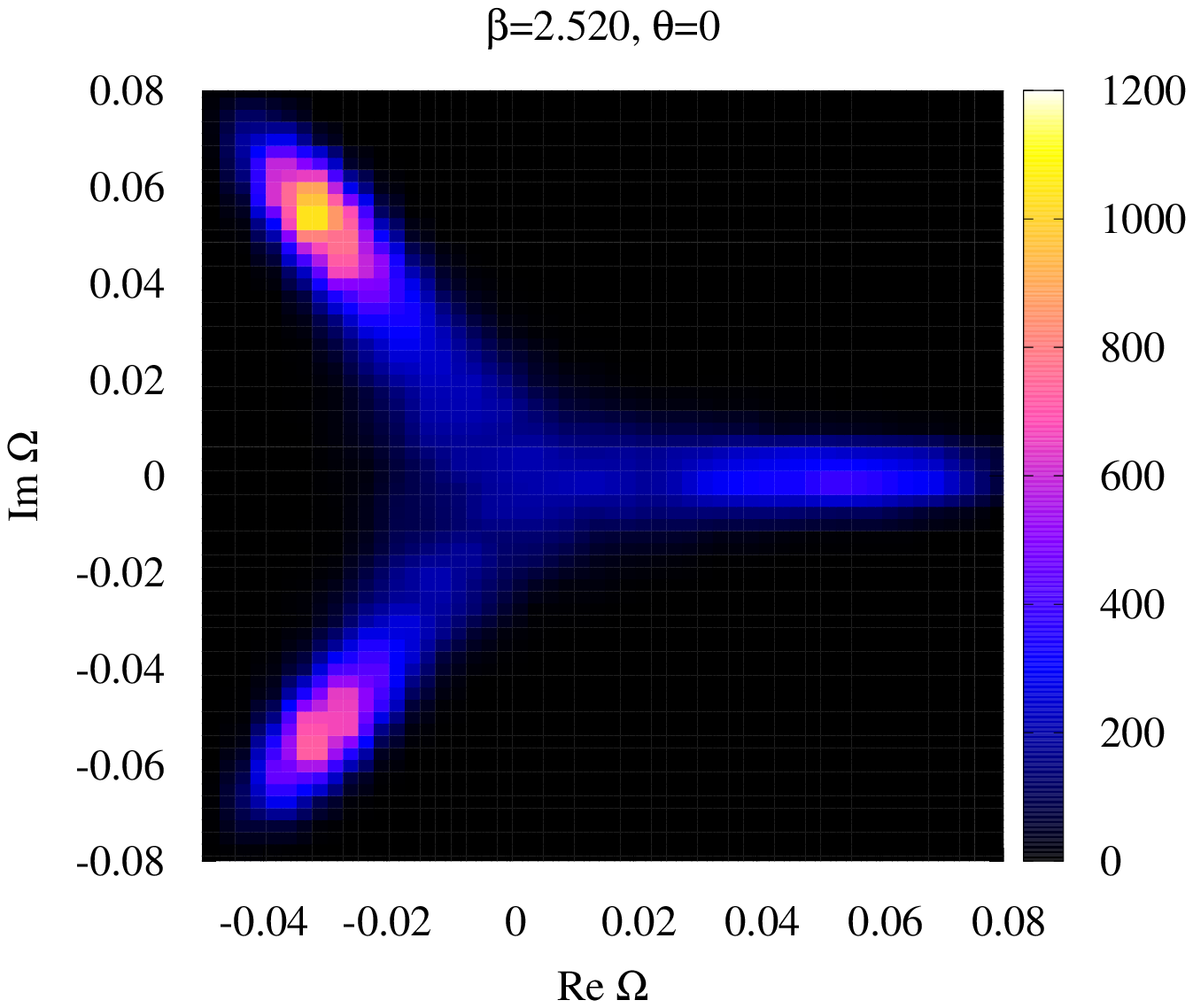}
  \end{tabular}  
 \end{center}
 \caption{Two dimensional histogram on the
 ${\rm Re}\,\Omega$-${\rm Im}\,\Omega$ plane.
 The results at $\beta=2.505$, 2.510, 2.515 and 2.520 are shown from
 left to right.
 }
 \label{fig:iws-poly-complex}
\end{figure}

Figure~\ref{fig:iws-poly-complex} shows examples with $N_o=2$,
the histogram of the real and imaginary part of the Polyakov loop at
$\th=0$, $p({\rm Re}\,\Omega,{\rm Im}\,\Omega;\beta,0)$, where it is
seen that the center symmetry is gradually broken with $\beta$.

\begin{figure}[tbp]
 \begin{center}
  \begin{tabular}{cc}
  \hspace*{-2ex}
  \includegraphics[width=0.37 \textwidth]
  {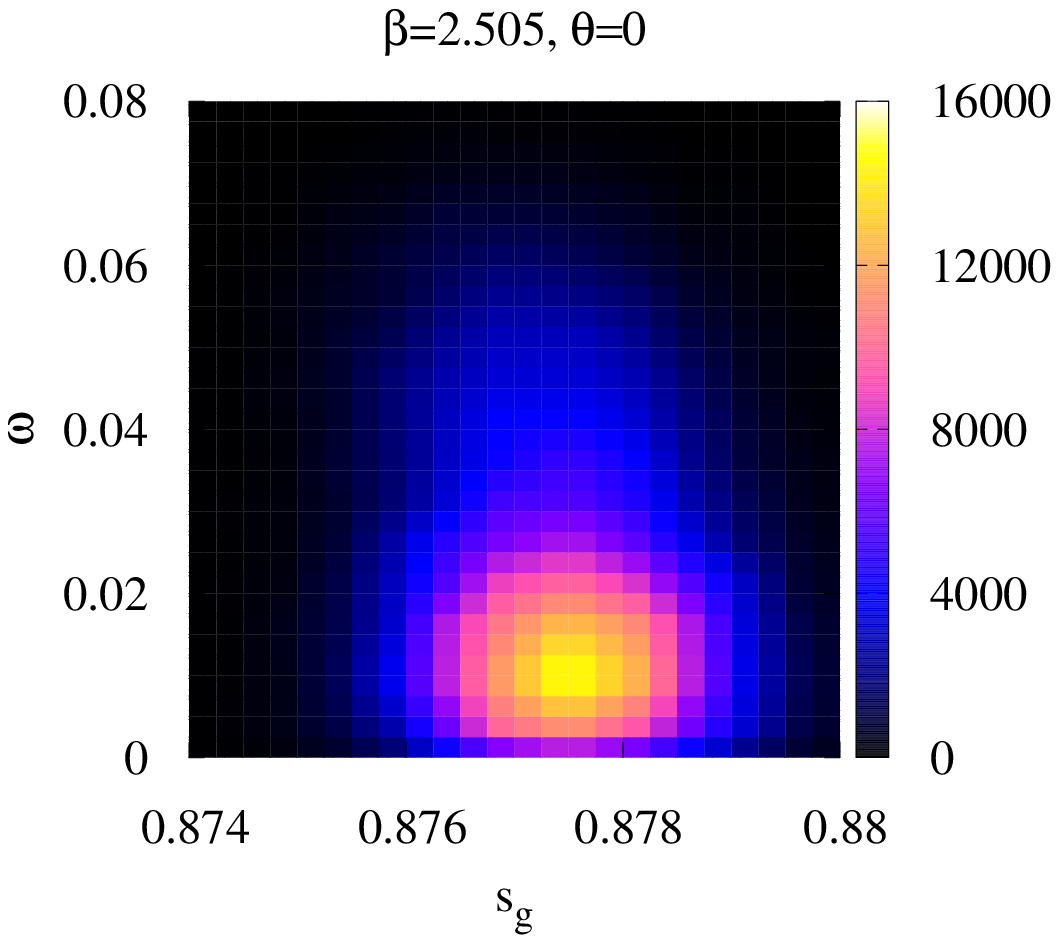} &
  \hspace*{-2ex}
  \includegraphics[width=0.37 \textwidth]
  {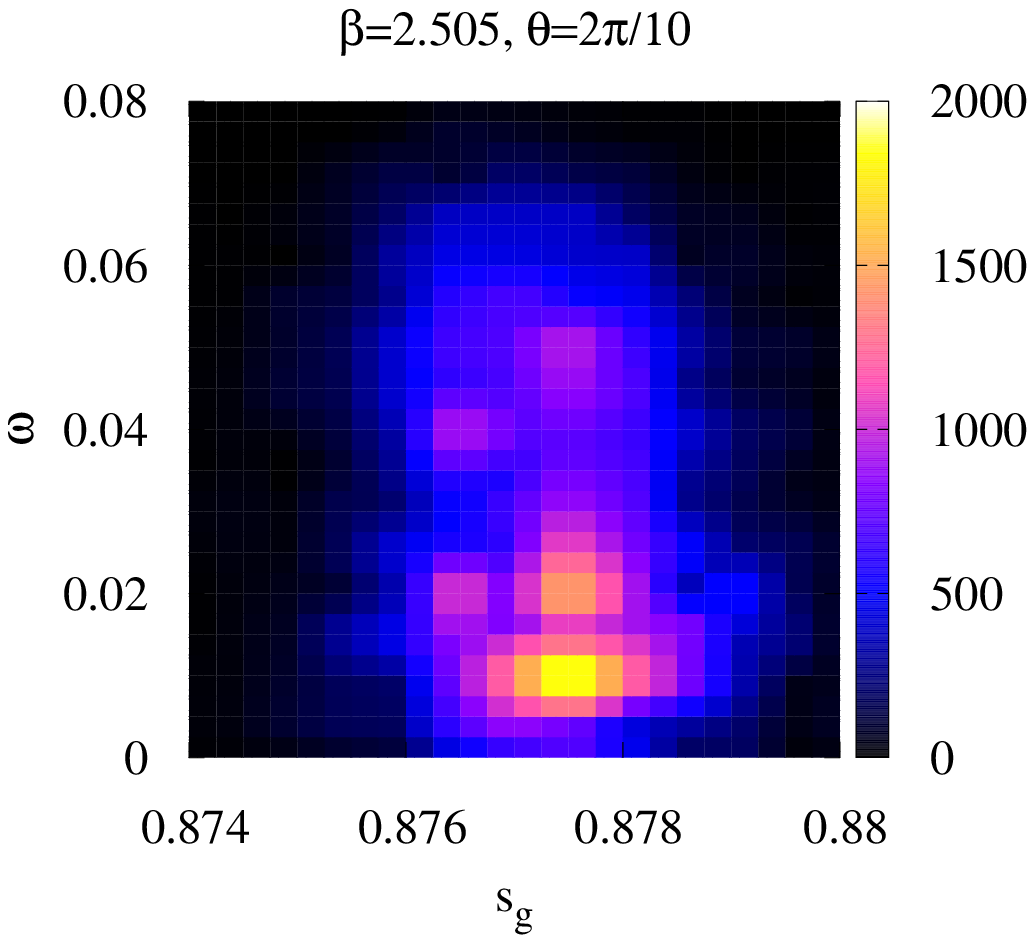} \\
  \hspace*{-2ex}
  \includegraphics[width=0.37 \textwidth]
  {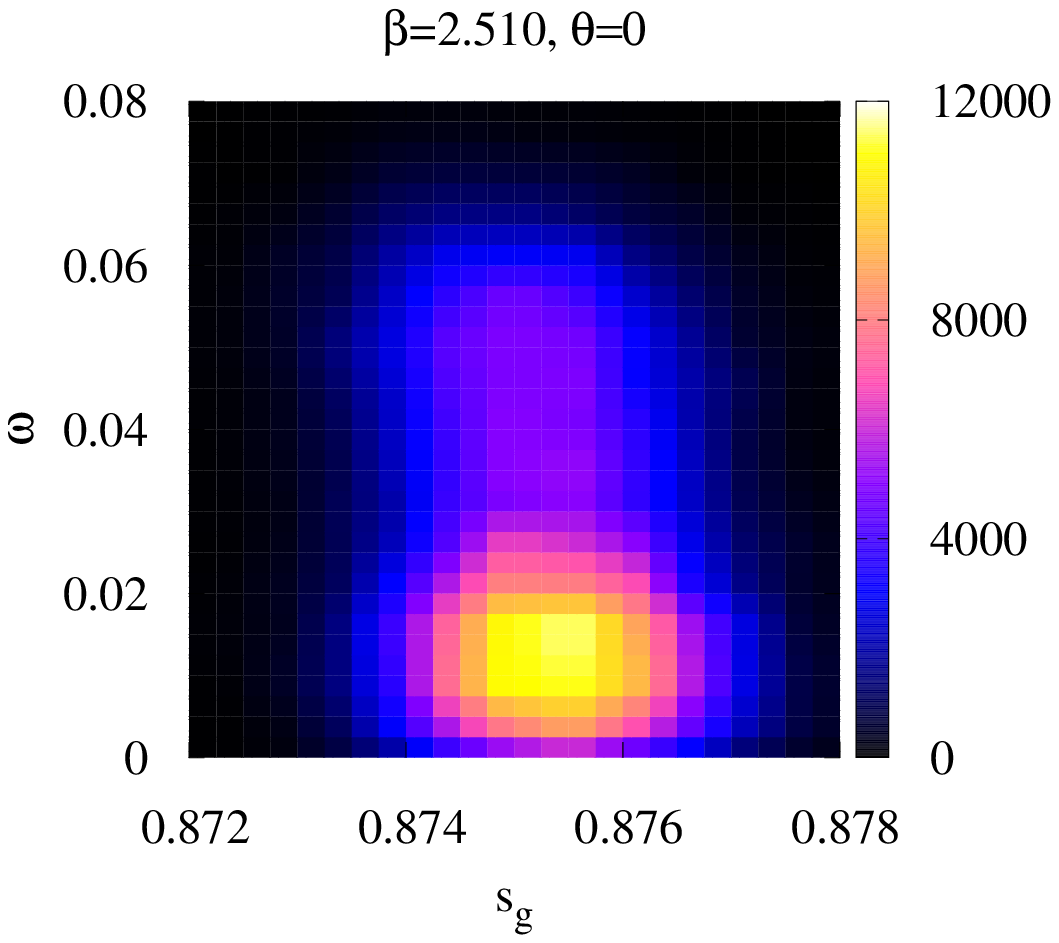} &
  \hspace*{-2ex}
  \includegraphics[width=0.37 \textwidth]
  {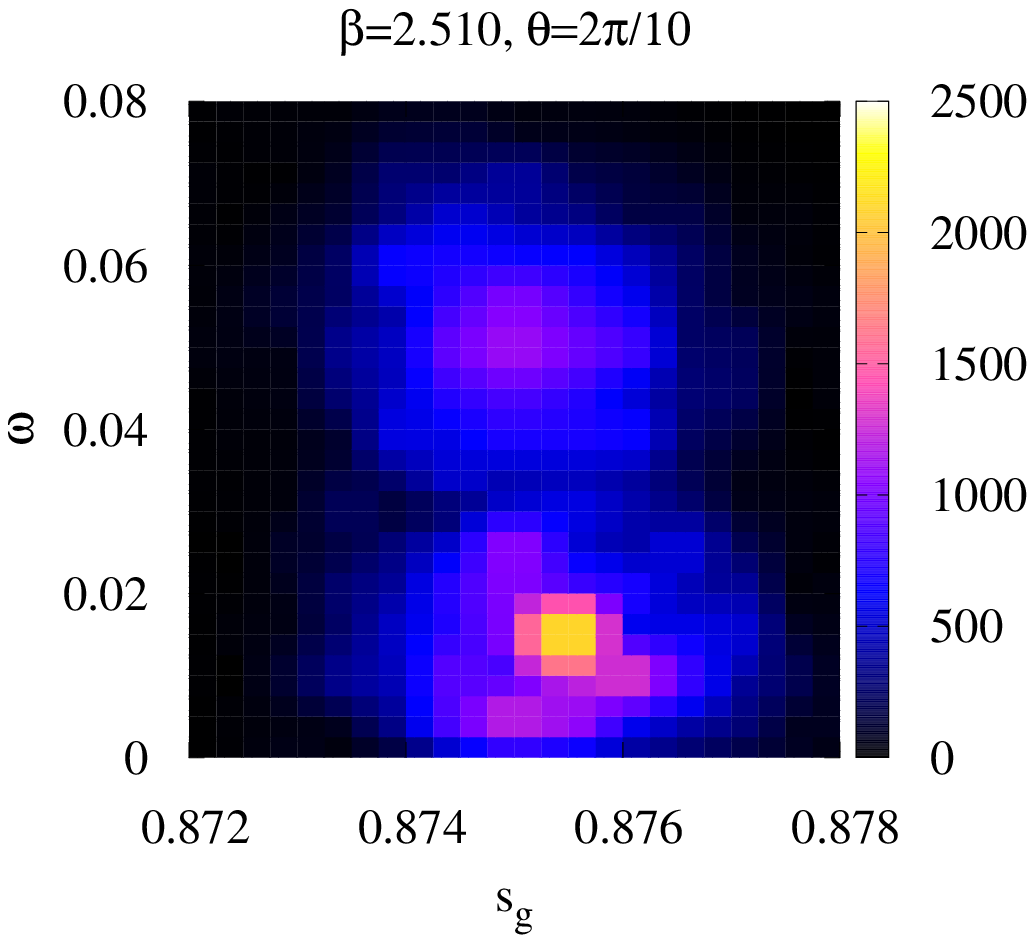} \\
  \hspace*{-2ex}
  \includegraphics[width=0.37 \textwidth]
  {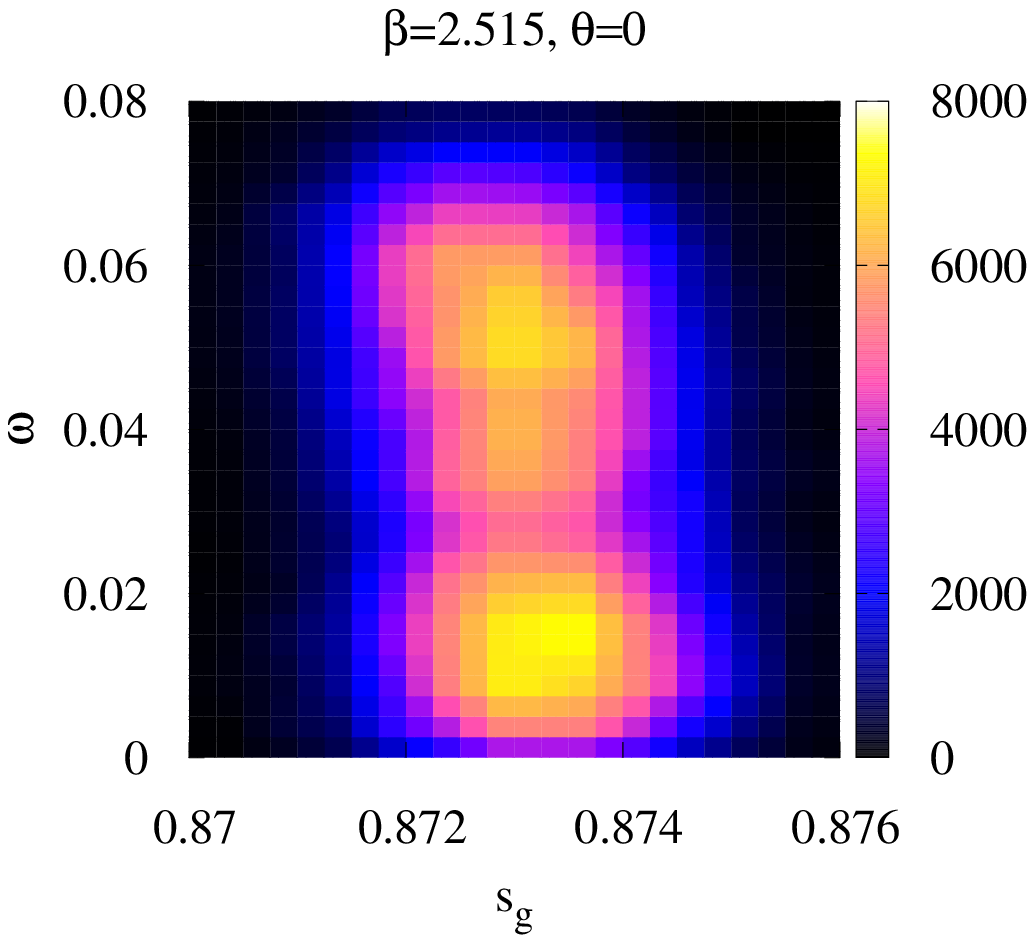} &
  \hspace*{-2ex}
  \includegraphics[width=0.37 \textwidth]
  {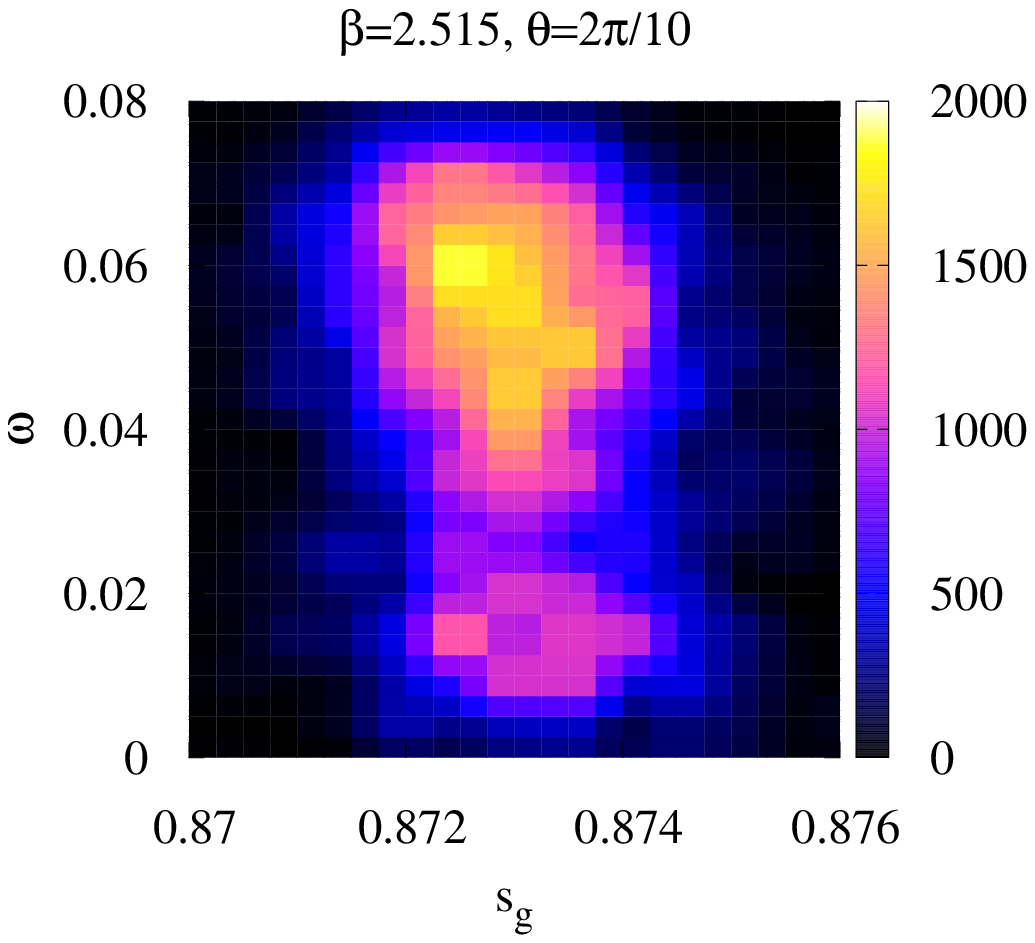} \\
  \hspace*{-2ex}
  \includegraphics[width=0.37 \textwidth]
  {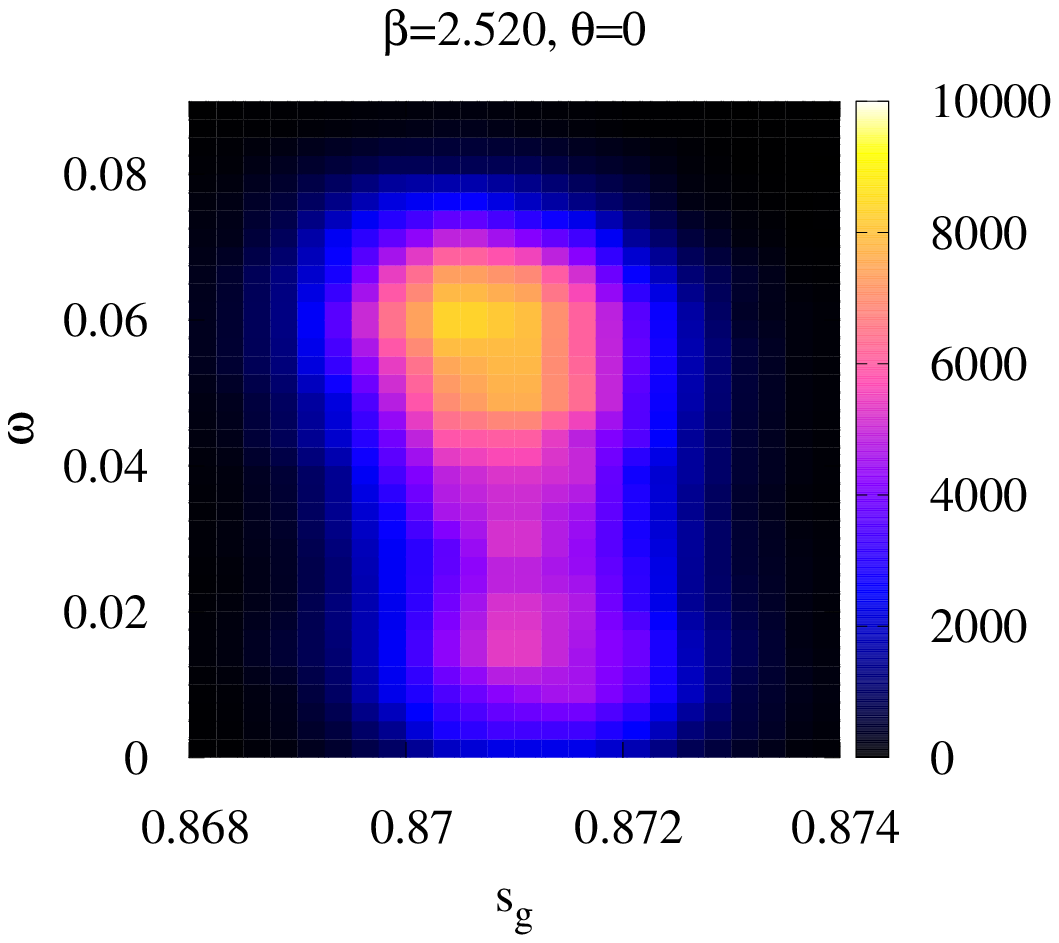} &
  \hspace*{-2ex}
  \includegraphics[width=0.37 \textwidth]
  {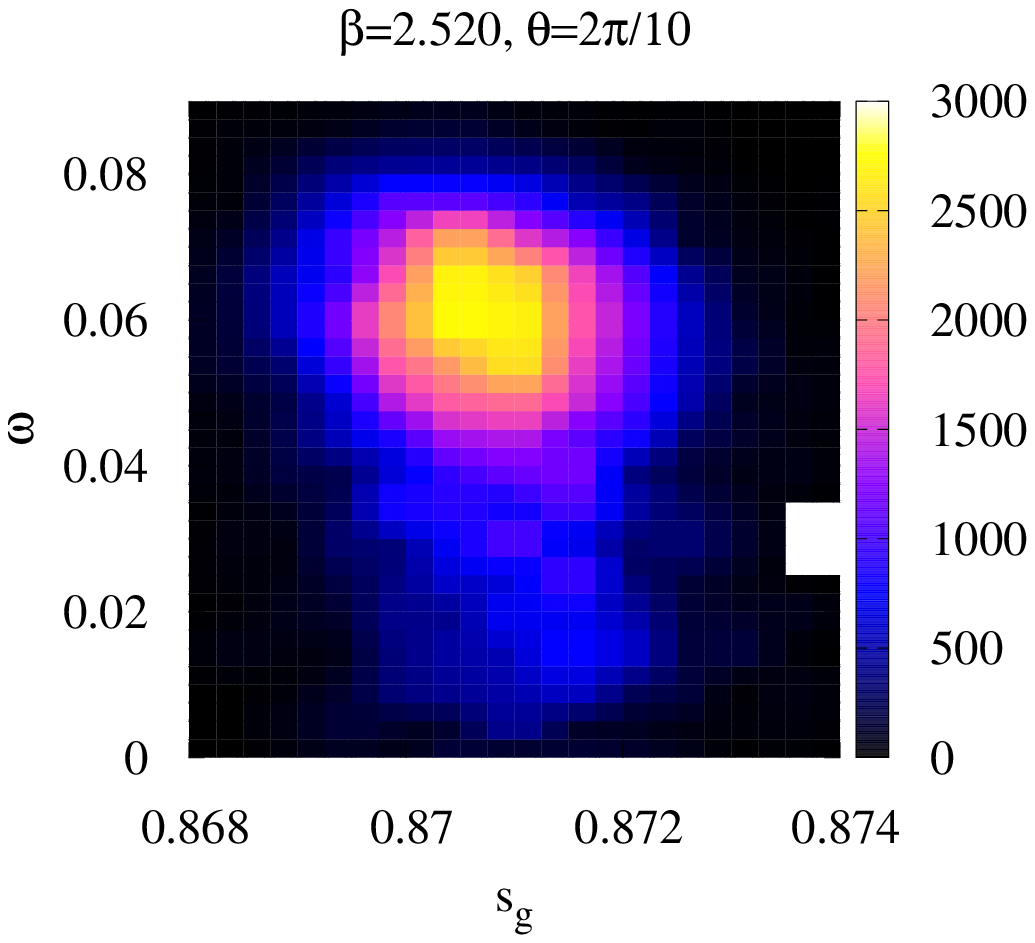}
  \end{tabular}
 \end{center}
 \caption{Two dimensional histogram on the $s_g$-$\omega$ plane at
 $\beta=2.505$ 2.510, 2.515, 2.520 from top to bottom.
 }
 \label{fig:iws-sg-omega}
\end{figure}

Another example with $N_o=2$ shown in Fig~\ref{fig:iws-sg-omega} is the
histogram for $s_g$ and $\omega$, $p(s_g,\omega;\beta,\th)$, at $\th=0$
and $2\pi/10$.
These histograms are used in the following analyses and the intervals of
histogram are set to 0.0005 for $s_g$ and 0.005 for $\omega$.
One or two peaks are observed around $\omega\sim 0.01$ and/or $0.05$, where
the former gets lower and the latter higher with $\th$.
In the plot for $\th=0$ and $\beta=2.515$, the two peaks are located at
almost the same value of $s_g$ and hence it would be difficult to
identify $\beta_c$ only by looking at the histogram for $s_g$, which is
contrary to the case using the Wilson plaquette gauge
action~\cite{Saito:2013vja}.

\begin{figure}[tb]
 \begin{center}
  \vspace*{-0ex}
  \begin{tabular}{cc}
  \hspace*{-2ex}
  \includegraphics[width=0.5 \textwidth]
  {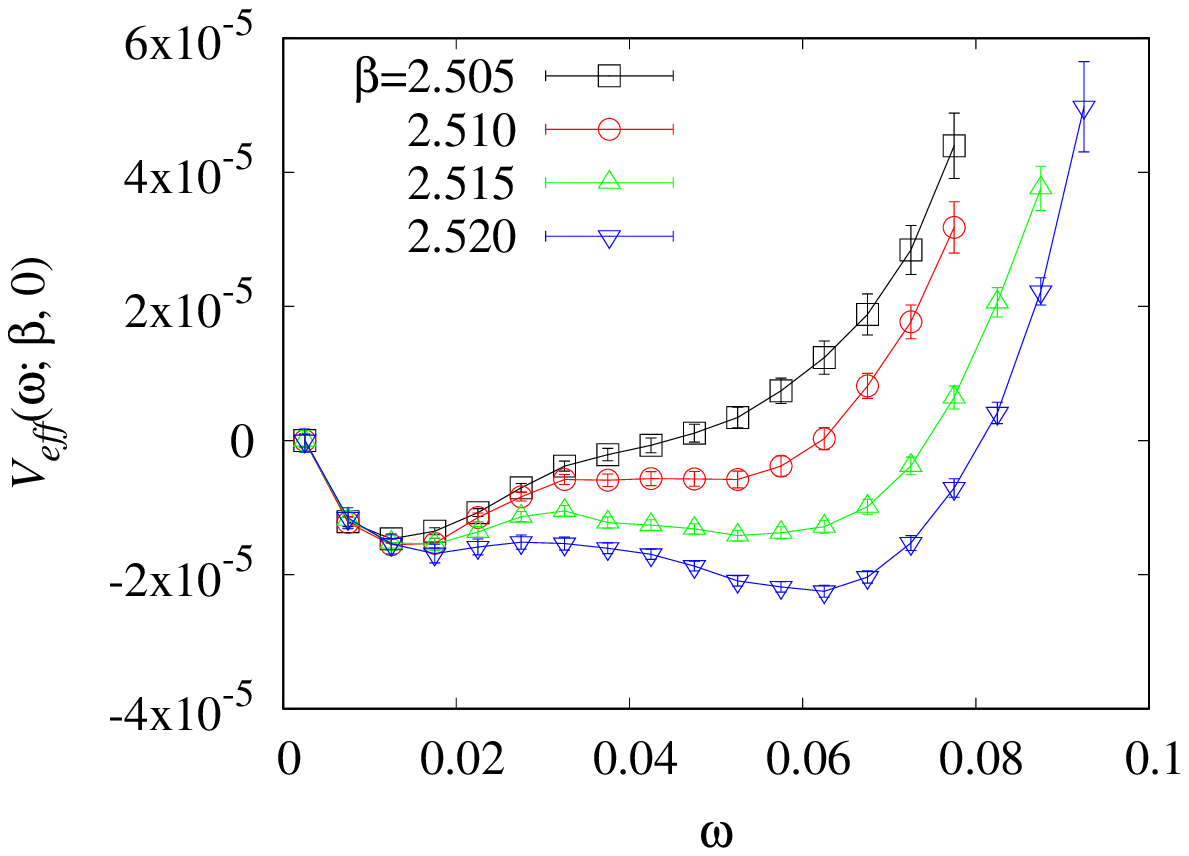} &
  \hspace*{-2ex}
  \includegraphics[width=0.5 \textwidth]
  {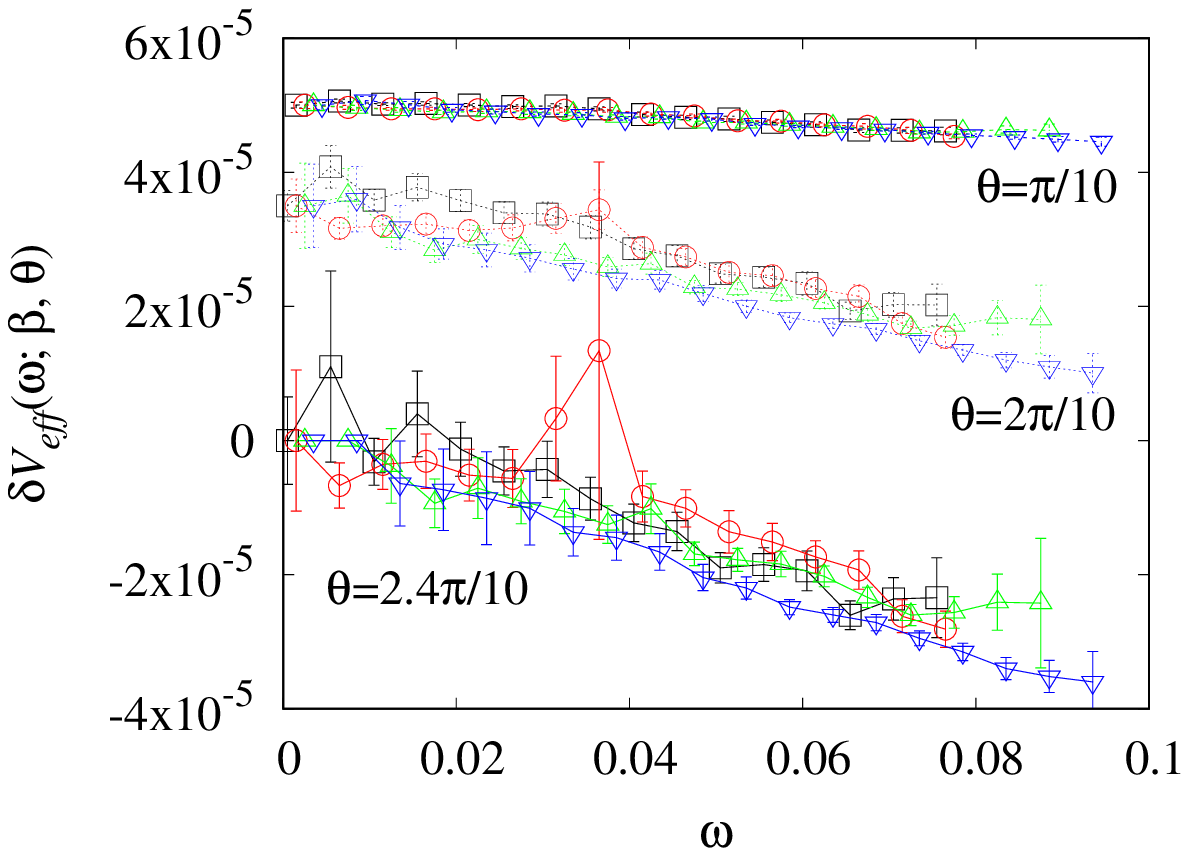}
  \end{tabular}
 \end{center}
 \caption{
 $V_\mathrm{eff}(\omega;\beta,0)$ (left) obtained at four $\beta$ values
 and $\delta V_\mathrm{eff}(\omega;\beta,\th)$ (right) for four $\beta$
 and three different $\th$.
 In both plots, the same symbols are used and the data are shifted in
 the vertical direction for visibility.
 }
 \label{fig:vpol}
\end{figure}

Using a histogram like above, we can define the constraint effective
potential by
\begin{align}
    V_{\rm eff}(o_1,\cdots,o_{N_o};\beta,\th) 
= - \frac{1}{\Nsite}\ln p(o_1,\cdots,o_{N_o};\beta,\th) \ ,
\label{eq:definition-v}
\end{align}
which coincides with the ordinary effective potential in the infinite
volume limit~\cite{ORaifeartaigh:1986axd}.
We investigate the constraint effective potential for $\homg$,
\begin{align}
     V_{\rm eff}(\omega;\beta,\th)
&= - \frac{1}{\Nsite}\ln p(\omega;\beta,\th)
\label{eq:vpol}
\ .
\end{align}
We can immediately calculate $V_{\rm eff}(\omega;\beta,\th)$ by using
the histograms at four values of $\beta$ following \eqref{eq:gene-hist}.
To show $V_{\rm eff}(\omega;\beta,\th)$, we separate it into the $\th=0$
contribution and the correction to that due to non-zero $\th$ defined by
\begin{align}
     \delta V_{\rm eff}(\omega;\beta,\th)
=& V_{\rm eff}(\omega;\beta,\th) - V_{\rm eff}(\omega;\beta,0)
=\ - \frac{1}{\Nsite}
     \ln {p(\omega;\beta,\th)\over p(\omega;\beta,0)}
\ .
\label{eq:vpol-correction}
\end{align}
$V_{\rm eff}$ and $\delta V_{\rm eff}$ are shown
in~Fig.~\ref{fig:vpol}.
It is seen that at $\th=0$ the global minimum transitions around
$\beta=2.515$.
The contribution of the nonzero $\th$ to the potential turns out to
depend on $\th$ but not on $\beta$ strongly.
It is also found that the nonzero $\th$ contributions are approximately
linear in $\omega$ with a negative slope being steeper with $\th$.
These observations immediately tell us that $\beta_c$ decreases with
$\th$.

\begin{figure}[tbp]
 \begin{center}
  \begin{tabular}{cccc}
  \hspace*{-2ex}
  \includegraphics[width=0.25 \textwidth]
  {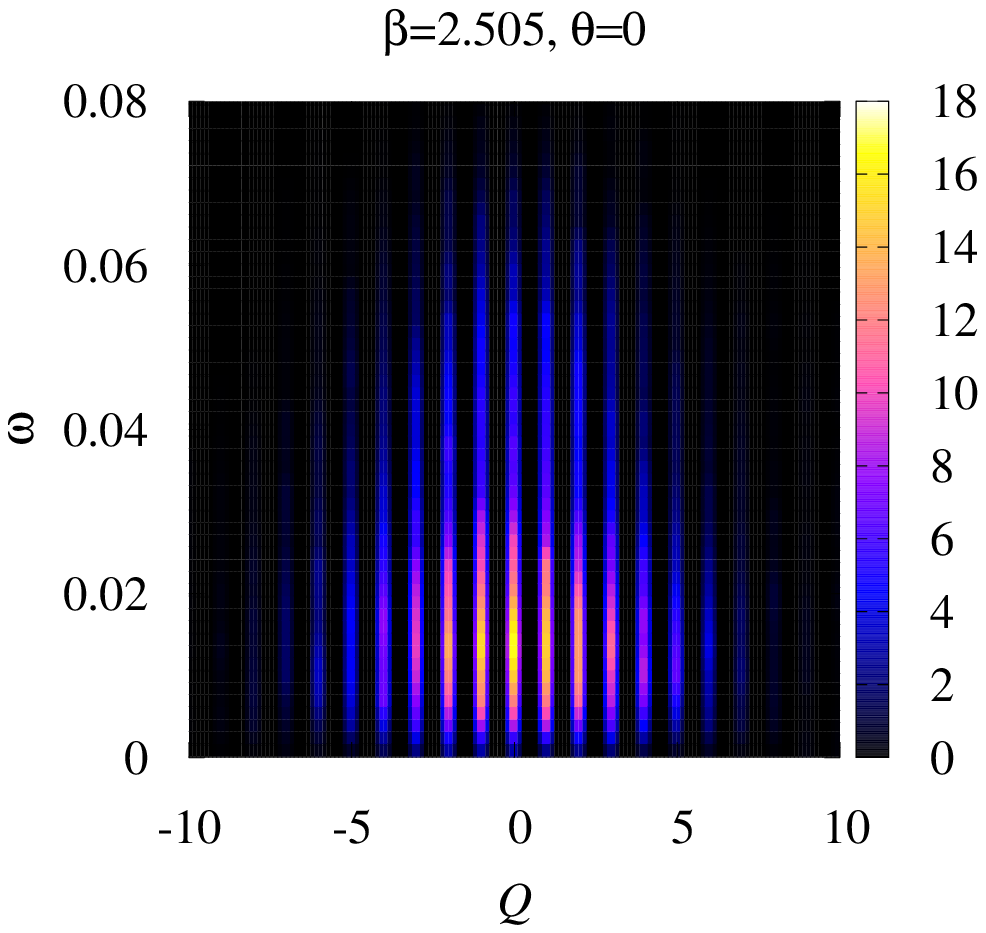} &
  \hspace*{-2ex}
  \includegraphics[width=0.25 \textwidth]
  {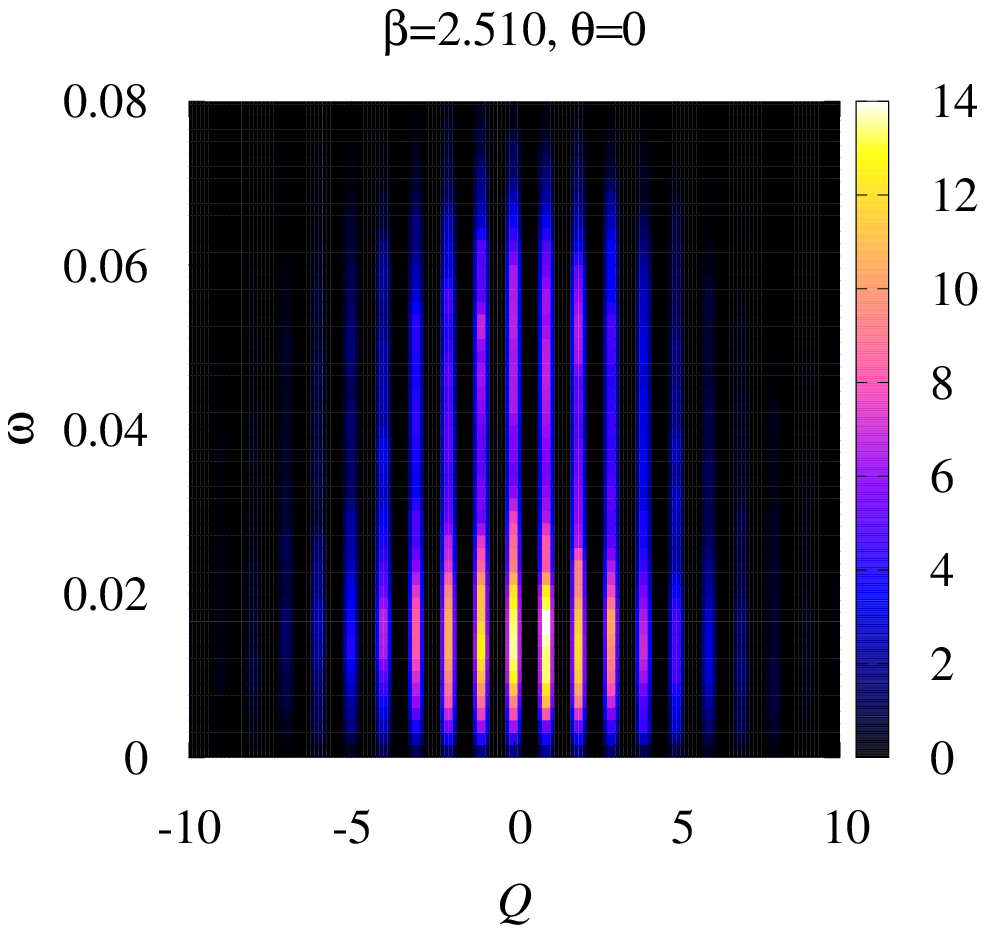} &
  \hspace*{-2ex}
  \includegraphics[width=0.25 \textwidth]
  {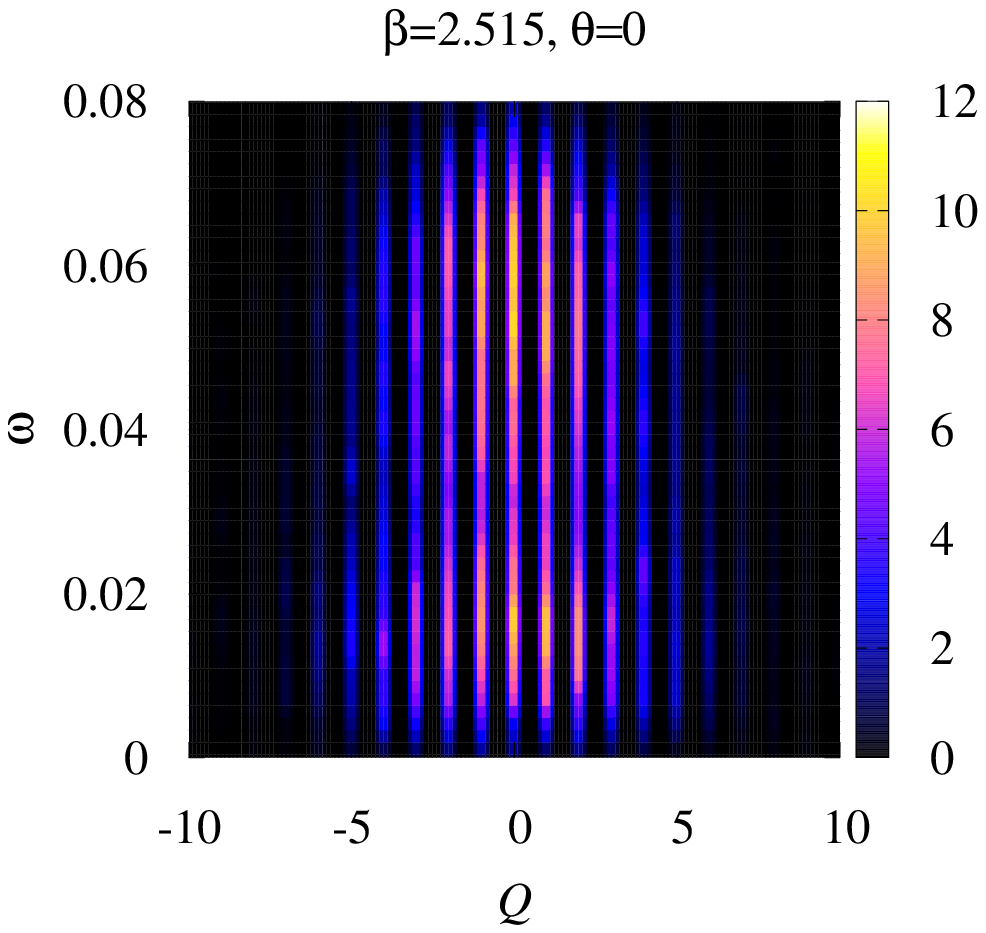} &
  \hspace{-2ex}
  \includegraphics[width=0.25 \textwidth]
  {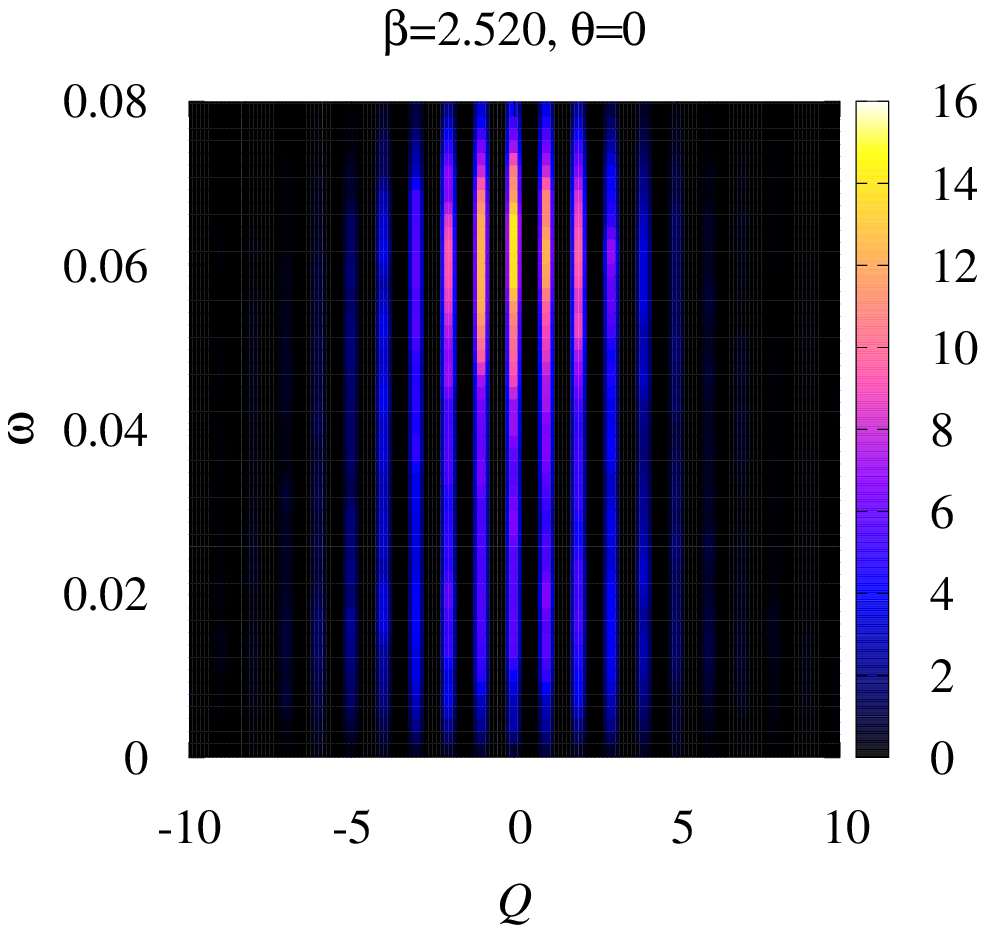}
  \end{tabular}  
 \end{center}
 \caption{Two dimensional histograms on the $Q$-$\omega$ plane at four
 values of $\beta$.
 }
 \label{fig:q-pol-hist}
\end{figure}

In order to explore the origin of the negative slope, we examine the
$Q$-$\omega$ histogram at $\th=0$, $p(Q,\omega;\beta,0)$, shown in
Fig.~\ref{fig:q-pol-hist}.
Roughly speaking, the histogram consists of two lumps located at
$\omega\sim 0.01$ and $\omega\sim 0.05$.
Importantly, the former spreads in the $Q$ direction more than the later.
Noticing that, up to a normalization,
\begin{align}
   p(\omega;\beta,\th)
&= \sum_{Q}p(Q,\omega;\beta,\th)
\sim \sum_{Q}\cos(\th Q)p(Q,\omega;\beta,0)
\ ,
\end{align}
the different distribution in $Q$ yields due to the cosine factor the
different suppression in $p(Q,\omega;\beta,\th)$ and then
$p(\omega;\beta,\th)$.
Since $p(Q,0.01;\beta,0)$ distributes wider in $Q$ than
$p(Q,0.05;\beta,0)$, $p(0.01;\beta,\th)$ is more suppressed.
It means that the local minimum in the potential around $\omega=0.01$
becomes relatively shallower compared with the one around $\omega=0.05$
when $\th$ increases from 0.
Therefore, the reason for the negative slope is attributed to the
different distribution of $Q$ in $p(Q,\omega;\beta,0)$ in a large and
small $\omega$ region.
The distribution of $Q$ is measured by the topological susceptibility,
$\chi=\langle\hQ^2\rangle_\beta/V$.
The relationship between the negative slope and the topological
susceptibilities in the low and high temperature phase becomes clear
when discussing the Clapeyron-Clausius equation later.

\subsection{interpolation in $\beta$}
\label{subsec:interpolating-beta}

We define $\beta_c$ by the value of $\beta$, at which two local minima
in the constraint effective potential degenerate.
As inferred from Fig.~\ref{fig:vpol}, $\beta_c$ varies with $\th$.
To determine $\beta_c$ at arbitrary $\th$ as precisely as possible, we
apply the multipoint reweighting
method~\cite{Ferrenberg:1989ui,Iwami:2015eba} to the histograms for
$\homg$, allowing us to interpolate them to desired values of $\beta$.

The details on how to interpolate the histogram in $\beta$ are described
below.
Recalling the reweighting technique, the histogram for $\hsg$ and
$\homg$ at $\beta$ and $\th $ can be written in terms of the expectation
values evaluated at $\beta_i$ and $\th=0$ as
\begin{align}
   p(s_g,\omega;\beta,\th)
=& {1\over Z(\beta,\th)}
   \int{\cal D}U\,\delta(\hsg-s_g)\delta(\homg-\omega)\,
   e^{-6\beta \Nsite \hsg-i\th\hQ}
\no\\
=& {\la \delta(\hsg-s_g)\delta(\homg-\omega) \cos(\th\hQ)\ra_{\beta_i}
    \over
    \la e^{-6(\beta-\beta_i)\Nsite (\hsg-s_g)} \cos(\th\hQ)\ra_{\beta_i}}
\ ,
\label{eq:pdf-beta}
\end{align}
where the numerator is given by \eqref{eq:gene-hist}.
By integrating \eqref{eq:pdf-beta} over $s_g$, one can obtain
$p(\omega;\beta,\th)$.
If there are more than one ensembles with different $\beta_i$'s, one can
calculate \eqref{eq:pdf-beta} on each of those ensembles and take an
average over them with suitable weights to determine the histogram at
$\beta$.

The simplest way of making the average would be
\begin{align}
   p(\omega;\beta,\th)
=\ {1 \over \left(\sum_i W_i\right)}
   \sum_i W_i\,\int\!\! ds_g\,
   {\la \delta(\hsg-s_g)\delta(\homg-\omega) \cos(\th\hQ)\ra_{\beta_i}
    \over
    \la e^{-6(\beta-\beta_i)\Nsite (\hsg-s_g)} \cos(\th\hQ)\ra_{\beta_i}}
\ ,
\label{eq:intplt-pdf-1}
\end{align}
with arbitrary weight $W_i$.
Following~\cite{Saito:2013vja}, as an alternative, we also calculate the
average by
\begin{align}
   p(\omega;\beta,\th)
=&\ \int\!\! ds_g\,
   {\sum_i W_i\,
    \la \delta(\hsg-s_g)\delta(\homg-\omega) \cos(\th\hQ)\ra_{\beta_i}
   \over
   \sum_i W_i
    \la e^{-6(\beta-\beta_i)\Nsite (\hsg-s_g)} \cos(\th\hQ)
    \ra_{\beta_i}}
\ ,
\label{eq:intplt-pdf-2}
\end{align}
which is obtained by rewriting \eqref{eq:pdf-beta} as
\begin{align}
    \la e^{-6(\beta-\beta_i)\Nsite (\hsg-s_g)} \cos(\th\hQ)\,
    \ra_{\beta_i}
    p(s_g,\omega;\beta  ,\th)
=&\ \la \delta(\hsg-s_g)\delta(\homg-\omega) \cos(\th\hQ)\ra_{\beta_i}
\ ,
\end{align}
and summing both sides over $i$ with a weight $W_i$.
We examined the above averages with two different weight factors
\begin{align}
W_i=&N(\beta_i)\ ,
\label{eq:weght-1}
\\
W_i=&N(\beta_i)\times
           \exp\{-(\beta-\beta_i)^2/\sigma_\beta^2\}\ ,
\label{eq:weght-2}
\end{align}
where $N(\beta_i)$ is the number of configurations generated at
$\beta_i$.
$\sigma_\beta$ essentially sets the effective range of $\beta_i$ to be
included in the average and is chosen to be $0.0025$ from the width of
the distribution of $s_g$ (the left panel of
Fig.~\ref{fig:iws-hist-sg-omg}).

It turns out that the combination of \eqref{eq:intplt-pdf-1} and
\eqref{eq:weght-1} is the noisiest among all and the others are
qualitatively similar.
In the following, we representatively show results with
\eqref{eq:intplt-pdf-2} and \eqref{eq:weght-1} as it does not contain
tunable parameters.

\begin{figure}[tbp]
 \begin{center}
  \vspace*{-0ex}
  \begin{tabular}{ccc}
  \hspace*{-1ex}
  \includegraphics[width=0.33 \textwidth]
  {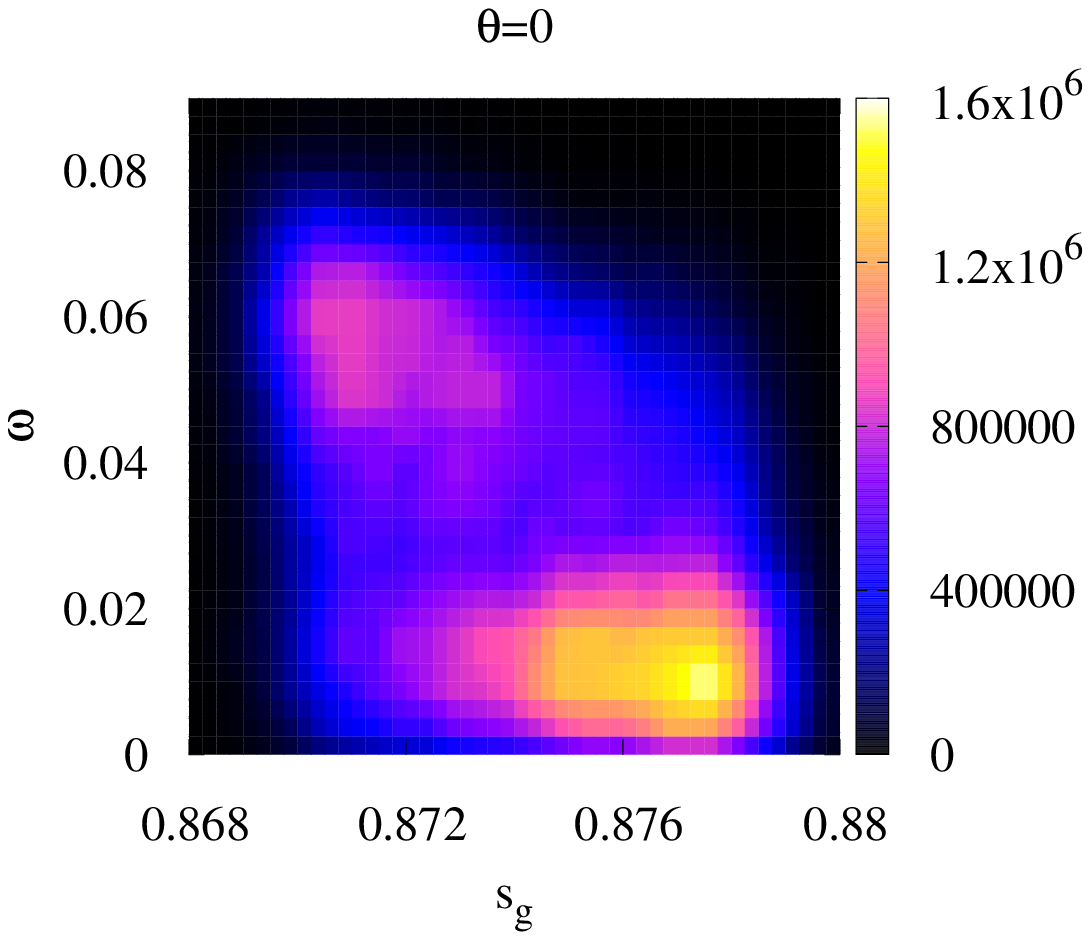} &
  \hspace*{-3ex}
  \includegraphics[width=0.33 \textwidth]
  {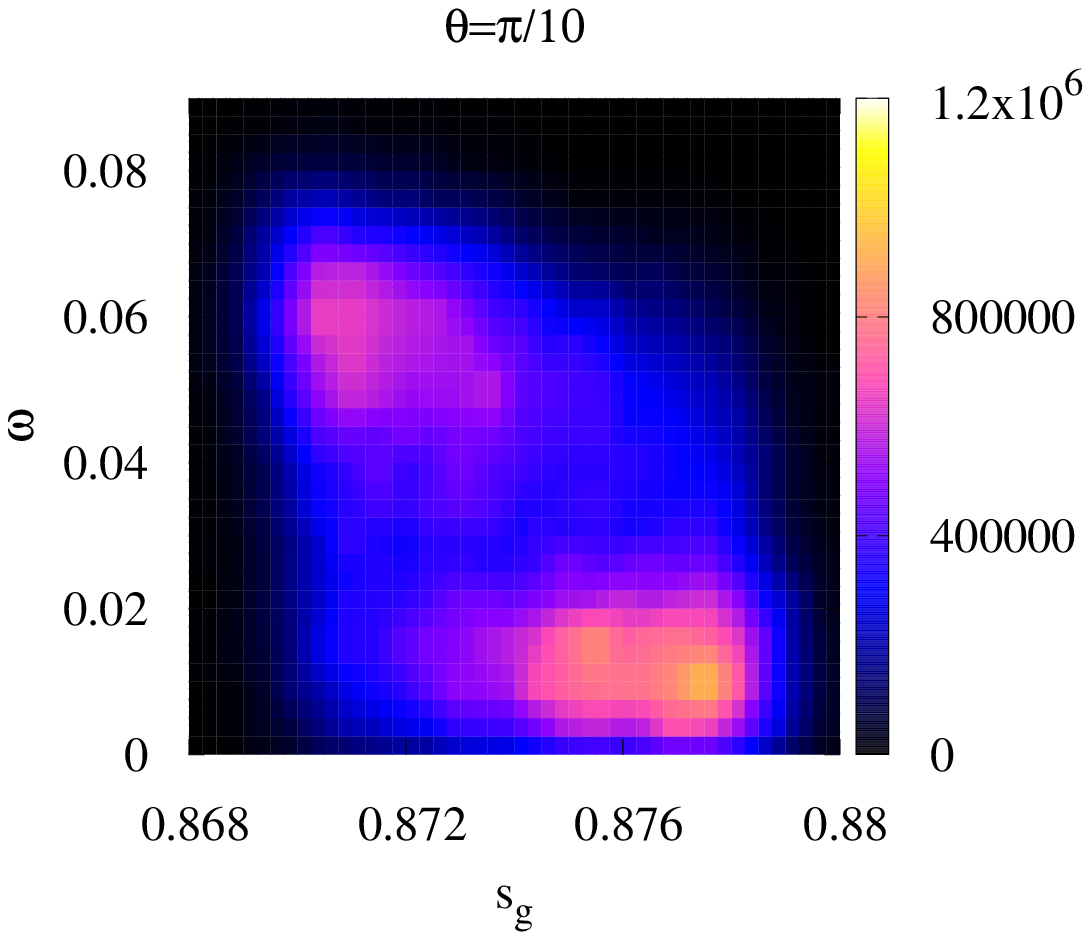} &
  \hspace*{-3ex}
  \includegraphics[width=0.33 \textwidth]
  {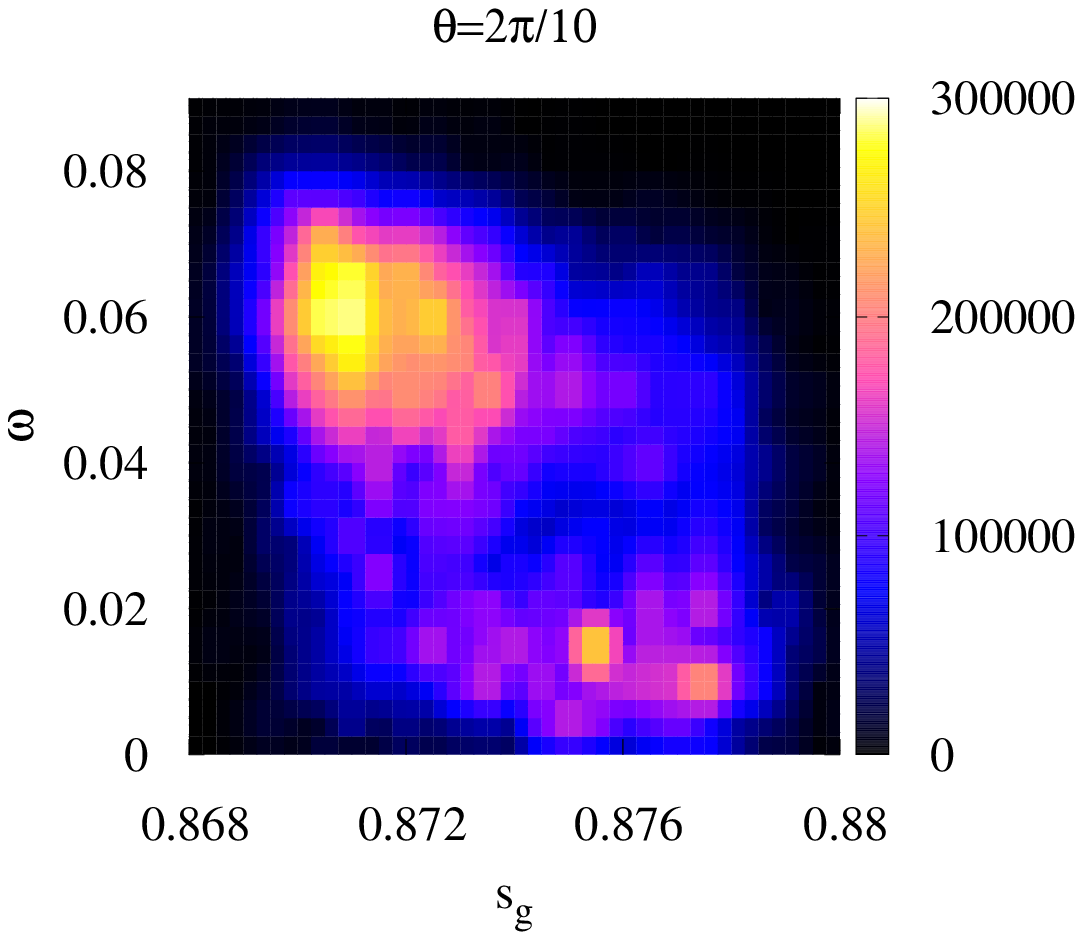}
  \end{tabular}
 \end{center}
 \caption{
  The numerator of the integrand in \eqref{eq:intplt-pdf-2} for $\th=0$,
 $\pi/10$ and $2\pi/10$ from left to right. 
 }
 \label{fig:nume_m2wa}
\end{figure}

\begin{figure}[tbp]
 \begin{center}
  \vspace*{-0ex}
  \begin{tabular}{ccc}
  \hspace*{-1ex}
  \includegraphics[width=0.33 \textwidth]
  {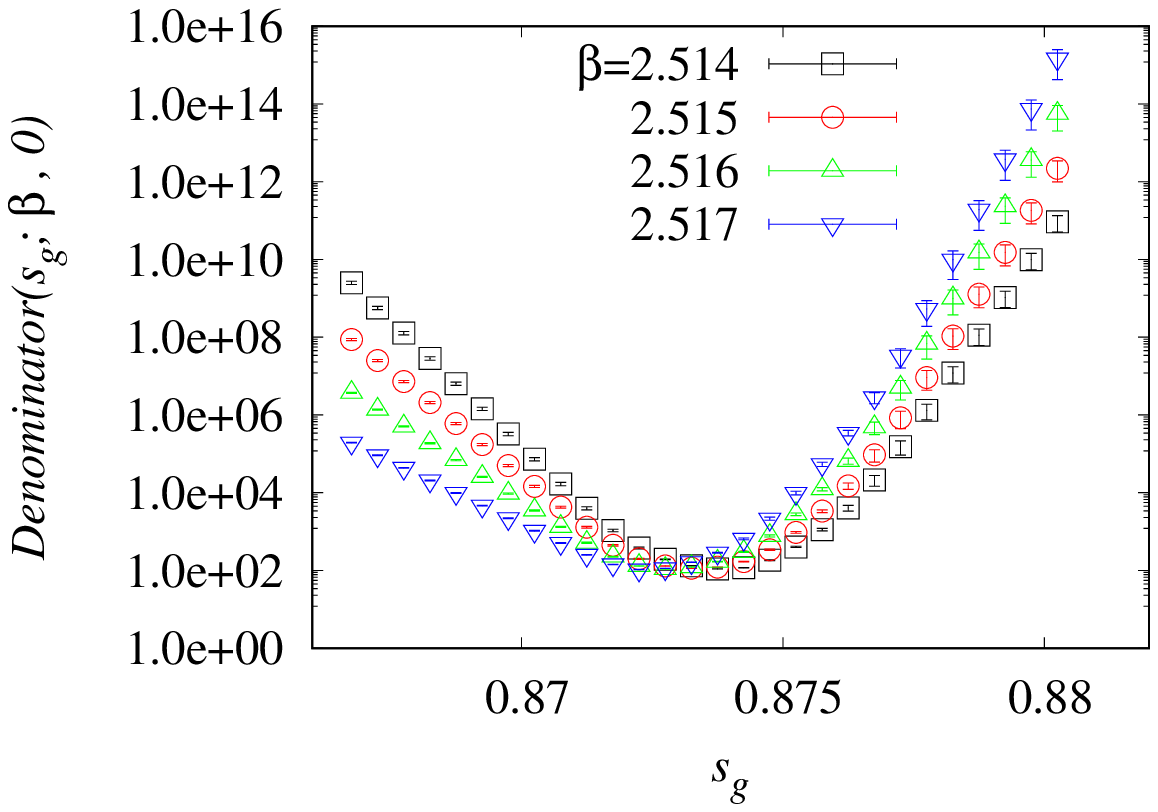} &
  \hspace*{-3ex}
  \includegraphics[width=0.33 \textwidth]
  {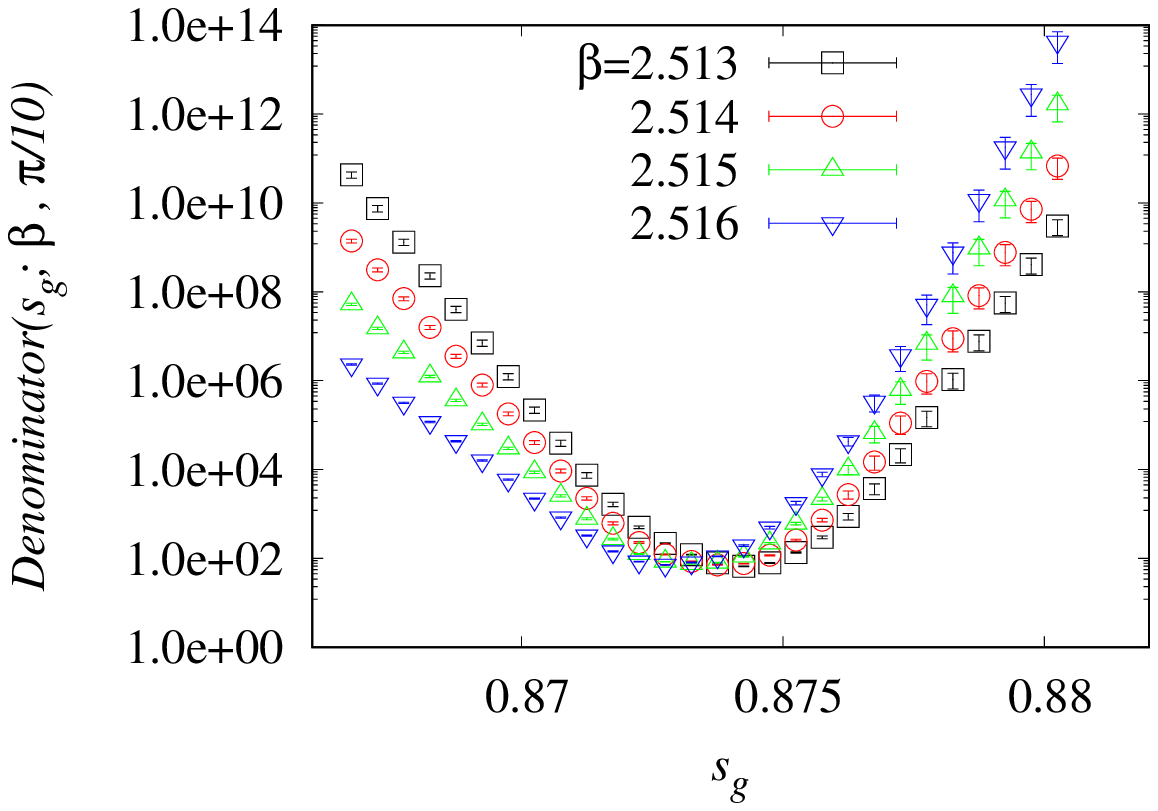} &
  \hspace*{-3ex}
  \includegraphics[width=0.33 \textwidth]
  {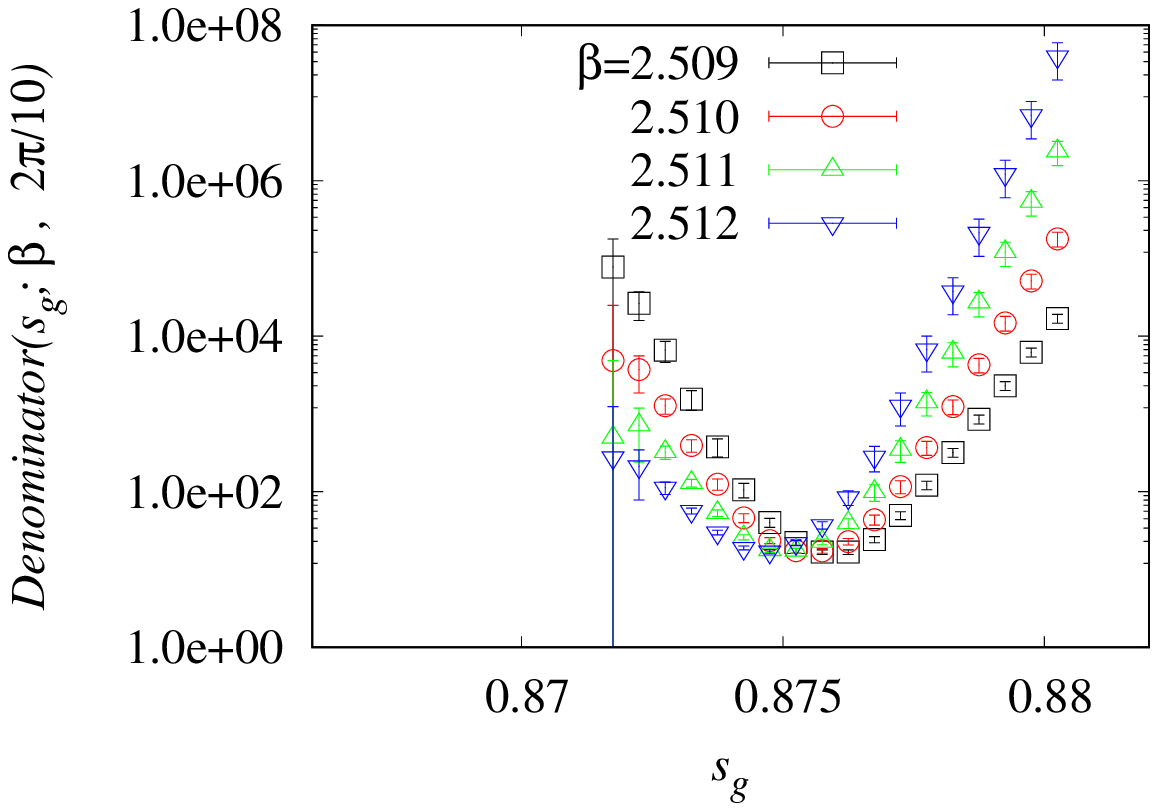}
  \end{tabular}
 \end{center}
 \caption{
  Some examples of the denominator of the integrand in
 \eqref{eq:intplt-pdf-2} for $\th=0$, $\pi/10$ and $2\pi/10$ from left
 to right. In the plot for $\th=2\pi/10$, the results below
 $s_g\sim  0.872$ are not shown because of too large statistical
 uncertainties.
 }
 \label{fig:deno_m2wa}
\end{figure}

\begin{figure}[ptb]
 \begin{center}
  \vspace*{-0ex}
  \begin{tabular}{ccc}
  \hspace*{-1ex}
  \includegraphics[width=0.33 \textwidth]
  {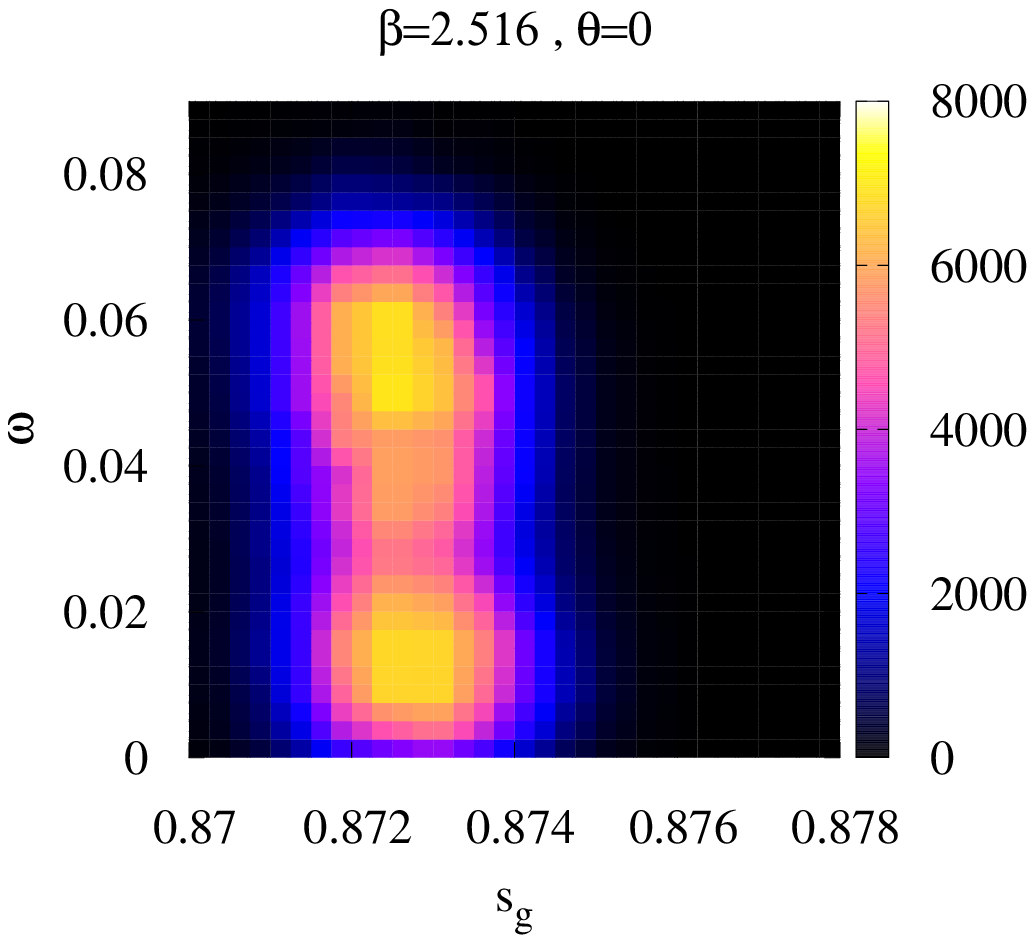} &
  \hspace*{-3ex}
  \includegraphics[width=0.33 \textwidth]
  {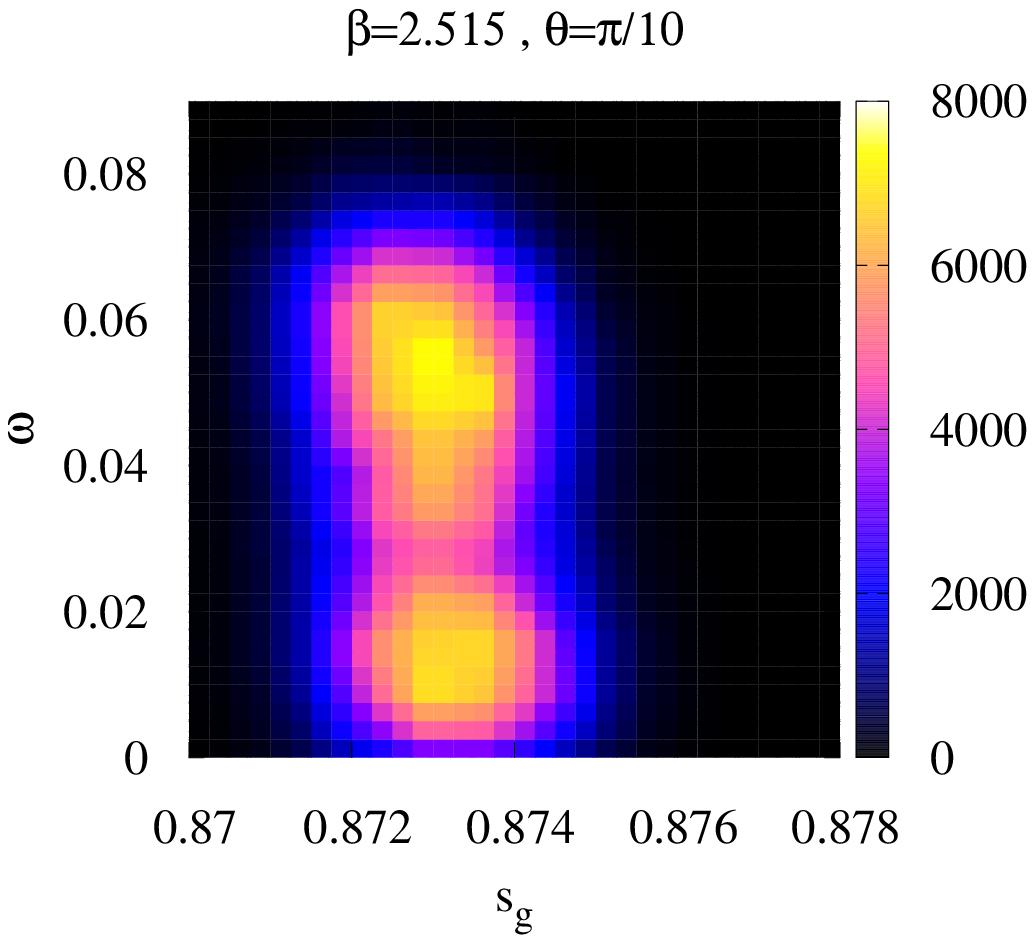} &
  \hspace*{-3ex}
  \includegraphics[width=0.33 \textwidth]
  {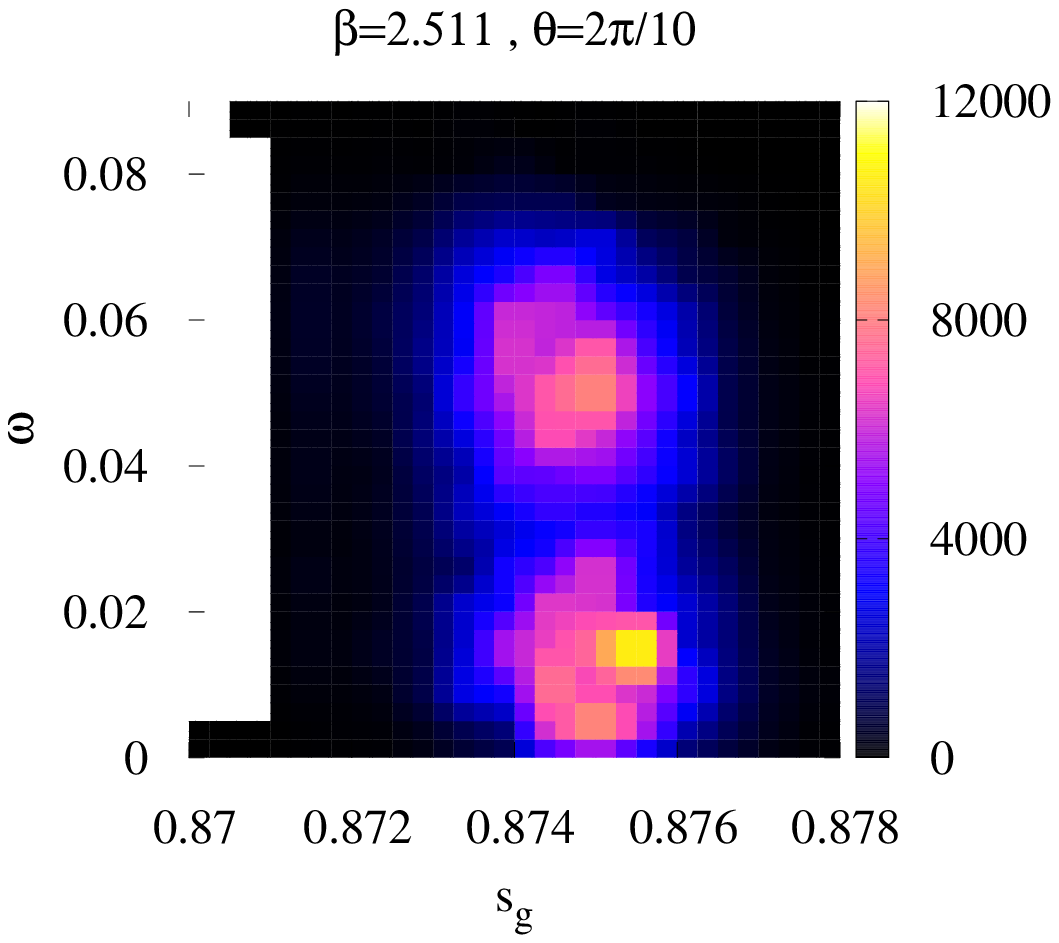}
  \end{tabular}
 \end{center}
 \caption{
  Examples of the resulting 2-d histogram, $p(s_g,\omega;\beta,\th)$,
 for $\th=0$, $\pi/10$ and $2\pi/10$ from left to right. 
 }
 \label{fig:2dhist_m2wa}
\end{figure}

Fig.~\ref{fig:nume_m2wa} shows the numerator of the integrand in
\eqref{eq:intplt-pdf-2}, which is the simple sum of the histogram
obtained at four ensembles with the weight \eqref{eq:weght-1}.
The denominator does not depend on $\omega$ and is the function of
$s_g$, $\beta$ and $\th$, some examples of which are shown in
Fig.~\ref{fig:deno_m2wa}.
As seen from the figure, the denominator takes a minimum at a certain
value of $s_g$, which depends on $\beta$.
Thus, around such $s_g$, the integrand in \eqref{eq:intplt-pdf-2} is
relatively enhanced.
Eventually, combining the numerator and denominator derives
$p(s_g,\omega;\beta,\th)$, some examples of which are shown in
Fig.~\ref{fig:2dhist_m2wa}.

$\beta$ in \eqref{eq:intplt-pdf-2} is arbitrary in principle, however
the interpolation does not always work in practice.
For example, if only a few configurations contribute to the enhanced
region of the histogram, the statistical error of that region becomes
large.
This necessarily happens if $\beta$ is not covered by the range of
$\beta_i$.
It is thus important to choose $\beta_i$'s over a suitable range and
with a small enough interval.
In this work, the interval of $\beta_i$ is chosen to be 0.005 such that
the distribution of $s_g$ at one of the ensembles well overlaps with
that at the neighboring $\beta_i$.

\section{Numerical results}
\label{sec:results}
\subsection{$\th$ dependence of $T_c$}

Integrating $p(s_g,\omega;\beta,\th)$ over $s_g$ and following
\eqref{eq:vpol}, the constraint effective potential for $\homg$ is
obtained.
Since for $\th>\pi/4$ the results are very noisy and $\beta_c(\th)$
appears to take the value out of the $\beta_i$ range, in the following
we restrict our analysis to $\th$ smaller than $\pi/4$.

\begin{figure}[t]
 \begin{center}
  \vspace*{-0ex}
  \begin{tabular}{cc}
  \hspace*{-2ex}
  \includegraphics[width=0.5 \textwidth]
  {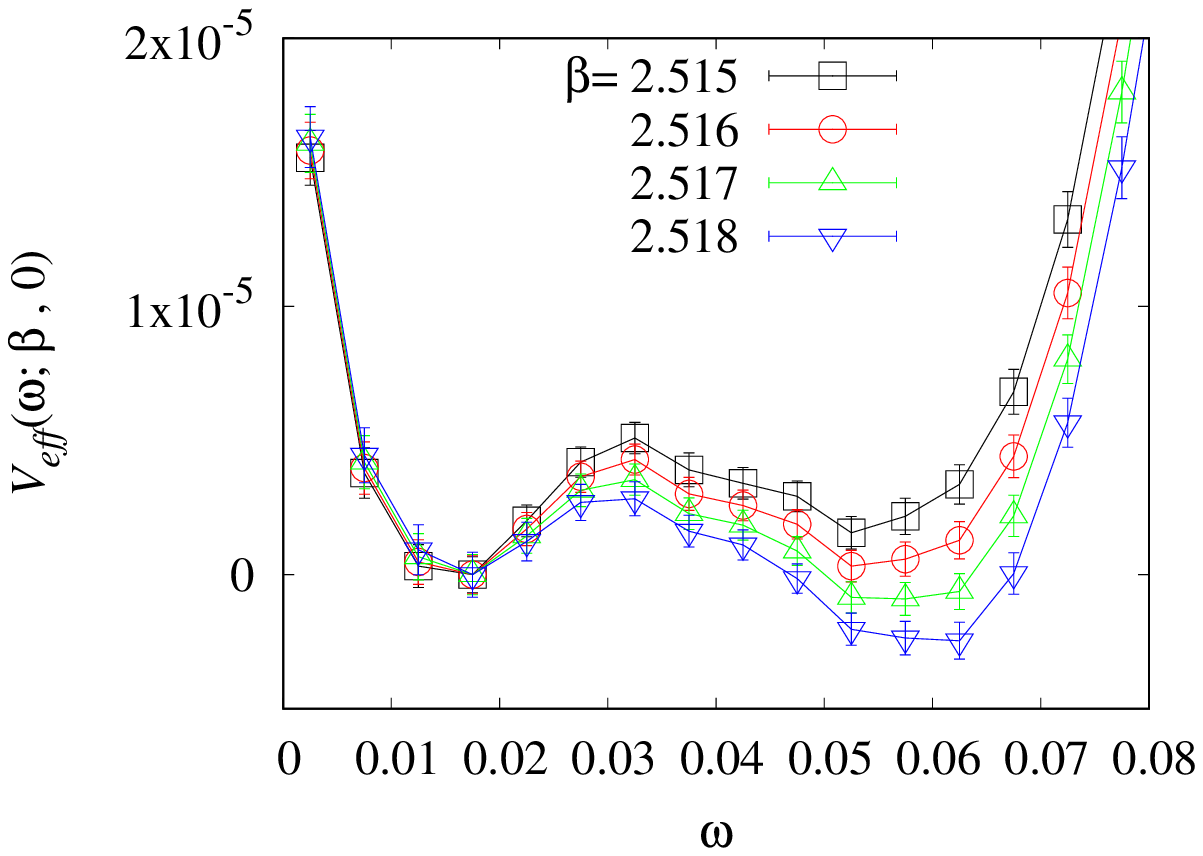} &
  \hspace*{-2ex}
  \includegraphics[width=0.5 \textwidth]
  {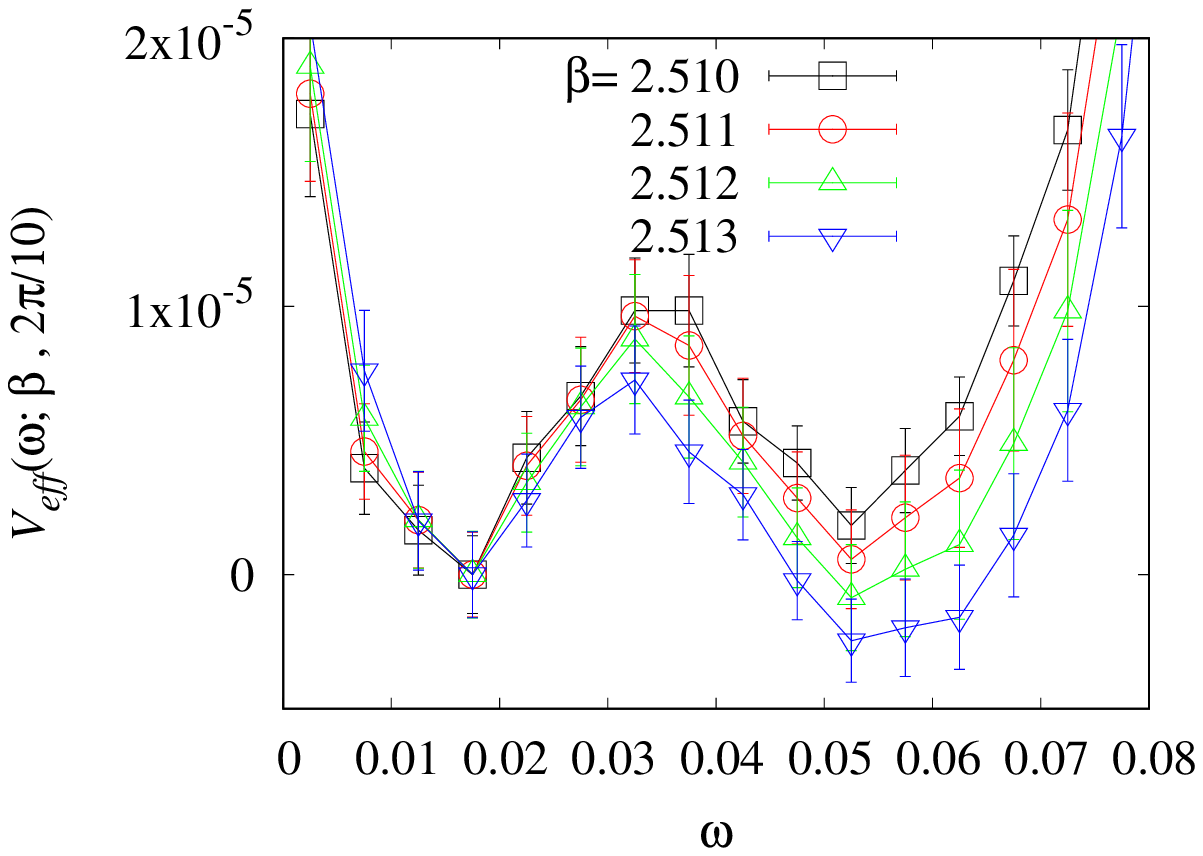}
  \end{tabular}
 \end{center}
 \caption{
  $V_{\rm eff}(\omega;\beta,\th)$ at $\th=0$ (left) and $2\pi/10$
 (right).
 }
 \label{fig:vpol_tune}
\end{figure}

Figure \ref{fig:vpol_tune} shows the potential at $\th=0$ and $2\pi/10$,
where the potential value at one of the minima around $\omega\sim 0.01$
is fixed to zero.
It is seen from Fig.~\ref{fig:vpol_tune} that the potential barrier
between two minima becomes higher with $\th$ in the lattice unit, which
suggests that the first order phase transition continues to be robust
beyond $\th\sim \pi/5$.

In addition to the robustness, we have also tried to see the $\th$
dependence of the strength by calculating the latent heat.
It looks consistent with constant, but the sizable statistical errors
prohibits us from leading definite conclusions.

We determine $\beta_c$ at each $\th$ by identifying the $\beta$ value at
which the difference between two minima vanishes.
\begin{figure}[tb]
 \begin{center}
  \begin{tabular}{c}
  \includegraphics[width=0.7 \textwidth]
  {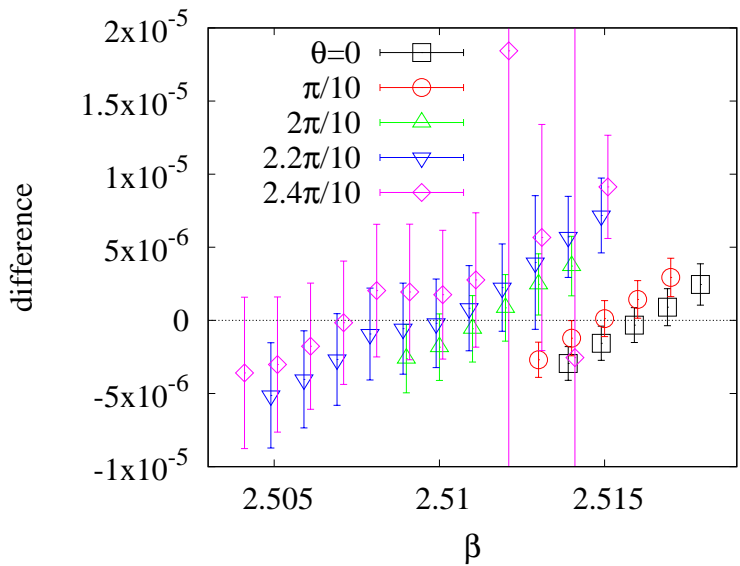}
  \end{tabular}
 \end{center}
 \caption{
  The difference of the energies between the two minima is shown as a
 function of $\beta$.
 }
 \label{fig:vpol_mindiff}
\end{figure}
The differences between two minima are plotted as a function of $\beta$
in Fig.~\ref{fig:vpol_mindiff} for $\th=0$, $\pi/10$, $2\pi/10$,
$2.2\pi/10$, $2.4\pi/10$.
The $\beta$ dependence of the difference turns out to be linear in the
region explored.
We thus fit them to a function linear in $\beta$ at each $\th$ and
obtain $\beta_c$.
$\beta_c(\th)$ is then altered to $T_c(\th)/T_c(0)$ using the relation
between the lattice spacing and $\beta$ determined at
$\th=0$ in~\cite{Okamoto:1999hi}.
The numerical results are tabulated in Tab.~\ref{tab:beta-crit}.
\begin{table}[ptb]
\begin{center}
 \begin{tabular}[t]{c|c|c}
  $\th$ & $\beta_c(\th)$ & $T_c(\th)/T_c(0)$\\
\hline
  0           & 2.5162( 9) & 1 \\
  $\pi/10$    & 2.5149( 9) & 0.9979( 6)\\
  $2\pi/10$   & 2.5112(15) & 0.9921(24)\\
  $2.2\pi/10$ & 2.5095(19) & 0.9894(31)\\
  $2.4\pi/10$ & 2.5074(32) & 0.9861(51)\\
 \end{tabular}
 \caption{The numerical results for $\beta_c(\th)$ and
 $T_c(\th)/T_c(0)$.}
 \label{tab:beta-crit}
\end{center}
\end{table}

\subsection{comparison with other results}
\label{subsec:comparison}

\begin{figure}[tb]
 \begin{center}
  \vspace*{-0ex}
  \includegraphics[width=0.7 \textwidth]
  {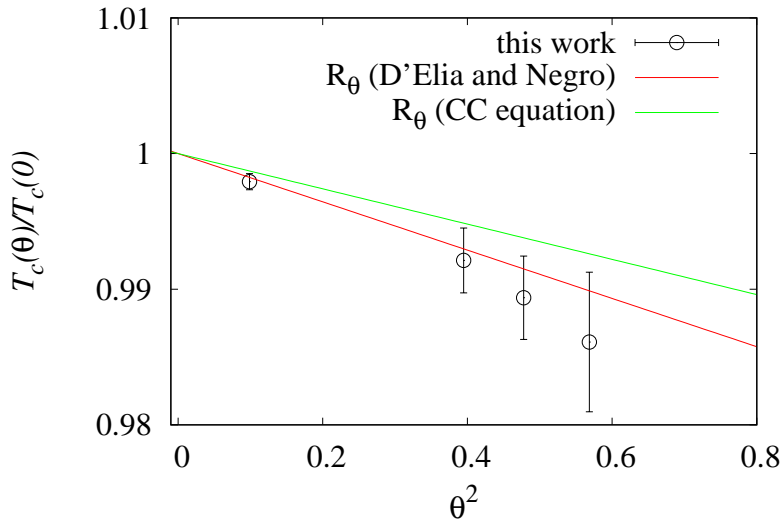}
 \caption{
  Comparison of our result for $T_c(\th)/T_c(0)$ with the result of
  \cite{DElia:2012pvq,DElia:2013uaf}.
  $R_\th$ obtained through the CC relation is also shown.
 }
 \label{fig:thetadep-tc}
 \end{center}
\end{figure}

D'Elia and Negro determined the $\th$ dependence of $T_c$ through the
Polyakov loop susceptibility~\cite{DElia:2012pvq,DElia:2013uaf}.
In their estimates, the analytic continuation from imaginary $\th$ and
the reweighting of real $\th$ give consistent results.
They parameterized the $\th$ dependence of $T_c$ as
\begin{align}
 {T_c(\th)\over T_c(0)}=1-R_\th \th^2+O(\th^4)\ ,
\label{eq:def-rth}
\end{align}
and derived $R_\th=0.0178(5)$ in the continuum limit.
The results obtained in this work and theirs are compared in
Fig.~\ref{fig:thetadep-tc}.
Although we have not yet taken the continuum limit, the reasonable
agreement is seen.

Another numerical test is possible by the use of the Clapeyron-Clausius
(CC) equation~\cite{DElia:2013uaf}.
The application of the equation to the present case is described in
Appendix~\ref{app:cc}, from which $R_\th$ is found to be
\begin{align}
R_\th
=&{\Delta\chi \over 2 \Delta\epsilon}\ ,
\label{eq:r-th}
\end{align}
where $\Delta\chi=\chi_L(T_c)- \chi_H(T_c)$ and $\chi_{L,H}(T_c)$
denote the topological susceptibility at the low and high temperature
phase at $T=T_c$, respectively.
For the latent heat, $\Delta\epsilon$, the precise values in the
continuum limit are available in~\cite{Shirogane:2020muc},
$\Delta\epsilon / T_c^4=$1.117(40) for $N_S/N_T=8$ and 1.349(38) for
$N_S/N_T=6$~\footnote{Recently, the precise value
$\Delta\epsilon / T_c^4=1.025(21)(27)$ is obtained in the infinite volume
and the continuum limits~\cite{Borsanyi:2022xml}}.
We extrapolate them using a function linear in $N_T/N_S$ to guess
$\Delta\epsilon/T_c^4\sim 1.813$ for $N_S/N_T=4$.

\begin{figure}[tb]
 \begin{center}
  \includegraphics[width=0.7 \textwidth]
  {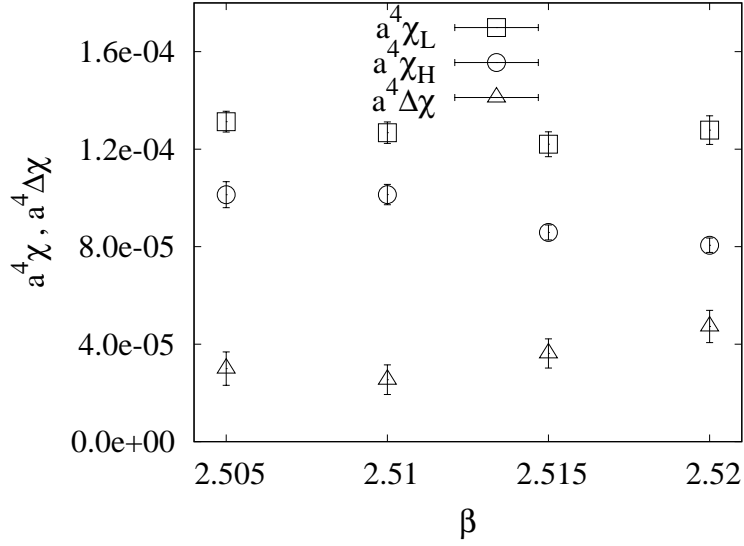}
 \caption{
  The difference of $a^4\chi$ in high and low temperature phases and
  their difference.
 }
 \label{fig:chi-diff}
 \end{center}
\end{figure}

For $\Delta\chi$, no data is available.
Thus, we try to estimate $\Delta\chi$ as follows.
We first divide the configurations at each $\beta_i$ into two groups, the
low and high temperature phases, by setting the threshold for $\omega$,
$\omega_{\rm th}$\footnote{For more sophisticated way to divides
configurations, see Ref.~\cite{Shirogane:2020muc}.}.
Taking $\omega_{\rm th}=0.030$, we calculate the topological
susceptibility in each group.
Figure~\ref{fig:chi-diff} shows the $\beta$ dependence of
$a^4\chi_L=\chi(\omega<\omega_{\rm th})$, 
$a^4\chi_H=\chi(\omega\ge \omega_{\rm th})$ and
$a^4\Delta\chi$.
Within the range of $\omega_{\rm th}$ we have studied
($0.028\le\omega_{\rm th}\le 0.032$), the variation of $a^4\Delta\chi$ is
negligibly small.
Since the $\beta$ dependence of $a^4\Delta\chi$ turns out to be mild, we
adopt $a^4\Delta\chi=3.6(6)\times 10^{-5}$ at $\beta=2.515$ as the gap
at the critical temperature at $\th=0$.

Substituting $\Delta\epsilon/T_c^4$ and $a^4\Delta\chi$ into
\eqref{eq:r-th} leads to
\begin{align}
R_\th
\sim 0.013 \ ,
\end{align}
which is reasonably consistent with the value by D'Elia and Negro if we
take into account the fact that the mass dimension of $\chi$ is four and
tends to be affected by large discretization effects.
$R_\th$ thus obtained is shown in Fig.~\ref{fig:thetadep-tc}.
Reasonable consistency with the numerical data for $T_c(\th)/T_c(0)$
implies that this way to estimate gaps at the first order transition
point can be used, at least, at qualitative level.

It is also possible to use our lattices to estimate
$\Delta\epsilon/T_c^4$ by recalling
$\Delta\epsilon / T^4=
6\,\Delta s_g N_T^4 a (d\beta / da)$~\cite{Saito:2011fs},
where $\Delta s_g=s_{gL}-s_{gH}$, the difference of the action density
in the low and high temperature phase.
Using the above divided configurations at $\beta=2.515$,
we obtain $\Delta s_g=2.66(12)\times 10^{-4}$, which gives
$R_\th\sim 0.0178$.

\subsection{free energy density across the transition curve}

The $\th$ dependence of the free energy density is another interesting
quantity to study because if it shows a cusp or any other non-analytic
behavior it signals a phase transition.
However, it is possible to observe such a behavior only after
accumulating enough statistics and taking the infinite volume limit.
Here, we consider how the free energy density is expected to behave as a
function of $\th$ when it crosses the $T_c(\th)$ curve and try to depict
it.
To this purpose, we choose the ensemble at $\beta=2.510$ as an
example and estimate the free energy density in the lattice unit,
$a^4 f(\th,T)$, defined by
\begin{align}
  a^4 f(\th,T)
= \lim_{\Nsite\to\infty}\frac{-1}{\Nsite}\ln \frac{Z(\beta,\th)}{Z(\beta,0)}
= \lim_{\Nsite\to\infty}\frac{-1}{\Nsite}
  \ln \,\la \cos\big(\theta\,\hQ\big)\ra_\beta
\ .
\label{eq:f-theta}
\end{align}
Since only single lattice size is available in this work, the infinite
volume limit is not taken in the following analysis.
Unfortunately, the simple implementation of \eqref{eq:f-theta} can not
detect possible signs of non-analytic behavior as the statistical error
and the finiteness of the volume obscure them.
Thus, we divide each configuration in the ensemble at $\beta=2.510$ and
determine $a^4 f(\th,T)$ in each phase through \eqref{eq:f-theta}.

\begin{figure}[tb]
 \begin{center}
  \includegraphics[width=0.7 \textwidth]
  {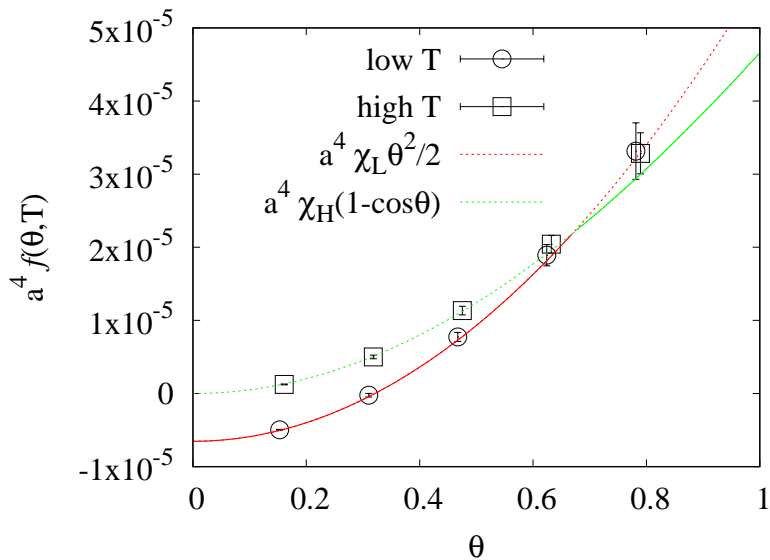}
 \caption{
  The $\th$ dependence of the free energy density.
  The numerical data are obtained at $\beta=2.510$.
  The expected behaviors in the low and high $T$ phases,
  $a^4\chi_L\th^2/2$ and $a^4\chi_H(1-\cos\th)$, are shown as a guide to
  eyes.
 }
 \label{fig:fed}
 \end{center}
\end{figure}

Figure~\ref{fig:fed} shows the $\th$ dependence of the free energy
densities.
The two curves represent $a^4f(\th,T)$ expected for $\th\ll 1$ in the two
phases, respectively, where $a^4 \chi_{L,H}$ shown in
Fig.~\ref{fig:chi-diff} are used.
According to Tab.~\ref{tab:beta-crit}, the free energy density at
$\beta_c\sim 2.510$ crosses the $T_c(\th)$ curve around 
$\th\sim 2.15\pi/10$.
Therefore, in Fig.~\ref{fig:fed} the data and the curve in the low $T$
phase are shifted in the vertical direction such that the two curves
meets there.
From the resulting free energy density, it is seen that the ground state
transitions from the low to the high $T$ phase as $\th$ increases from 0.

\section{zeros of partition function}
\label{sec:lee-yang-zeros}

When a system experiences a phase transition, the free energy often
shows non-analytic behavior, and correspondingly the partition function
takes zero in the infinite volume limit.
In finite volume, instead zeros appear in unphysical regions of a
parameter, known as the Lee-Yang zeros~\cite{Yang:1952be,Lee:1952ig}
or Fischer's zeros~\cite{Fischer:1965rna}.
The aim of this section is to see how the zeros of the partition
function associated with phase transitions in the $\th$-$T$ plane
appear.
First, following the discussion of Ref.~\cite{Ejiri:2005ts}, let us see
the Fischer's zeros, where $\beta$ is extended to a complex variable as
$\beta=\beta_R+i\beta_I$ and $\beta_R$ is fixed to the critical value
for a given real $\th$, $\beta_R=\beta_c=\beta_c(\th)$.
Consider the following ratio of the partition functions and write it in
terms of the histogram for $\hsg$ and $\homg$ as
\begin{align}
   {Z(\beta_c+i\beta_I,\th)\over Z(\beta_c,\th)}
&= {1 \over Z(\beta_c,\th)}
   \int\!\! d\omega\,ds_g\,\int{\cal D}U
   \delta(\homg-\omega)\,\delta(\hsg-s_g)
   e^{-6(\beta_c+i\beta_I)\Nsite\hsg-i\th\hQ}
\no\\
&= \int\!\! d\omega\,ds_g\,
   e^{-6i\beta_I\Nsite\,s_g}\,
   p(s_g,\omega;\beta_c,\th)
\label{eq:partition-f}
\ .
\end{align}
Assuming the volume to be large enough, when the first order phase
transition occurs, the integral of \eqref{eq:partition-f} is dominated
by two peaks in $p(s_g,\omega;\beta_c,\th_R)$ around $(s_{g1},\omega_1)$
and $(s_{g2},\omega_2)$ with an equal height, {\it i.e.}
$p(s_{g1},\omega_1;\beta_c,\th_R)=p(s_{g2},\omega_2;\beta_c,\th_R)$.
Up to an overall constant, \eqref{eq:partition-f} is simplified as
\begin{align}
&\ \int\!\! d\omega\,ds_g\,
   e^{-6i\beta_I\Nsite\,s_{g}}\,
   p(s_g,\omega,Q;\beta_c,0)
\no\\
&\approx
   \left(
     e^{-6i\beta_I\Nsite\,s_{g1}} + e^{-6i\beta_I\Nsite\,s_{g2}}
   \right)\,
   p(s_{g1},\omega_1;\beta_c,\th_R)
\no\\
&\approx
  2\,e^{-3i\beta_I\Nsite\,(s_{g1}+s_{g2})}
  \cos\big(3\,\beta_I\Nsite\,(s_{g1}-s_{g2})\big)\,
  p(s_{g1},\omega_1;\beta_c(\th_R),\th_R)
\ .
\end{align}
Therefore, $Z(\beta_c+i\beta_I,\th)$ in \eqref{eq:partition-f} vanishes at
\begin{align}
 (\beta_R,\ \beta_I)=
 \left(\beta_c,\ {\pm(2n+1)\pi\over 6\Nsite(s_{g1}-s_{g2})}\right)
\qquad {\rm with }\ n=0, 1, 2, \cdots
\ .
\end{align}
The zeros appear periodically in the imaginary direction, and their
interval is inversely proportional to the volume.
Therefore, the appearance of two peaks results in these typical signatures
for the first order phase transition.

Next, let us explore the Lee-Yang zeros in the complex $\th$ plane.
In the following, $\beta$ is real and $\th=\tR+i\tI$.
In this case, the Lee-Yang zeros are found in a specific case.
Consider theories in which the histogram of the topological charge at
$\th=0$ follows the Gaussian distribution,
\begin{align}
   p(Q;\beta,0)
\sim e^{-\frac{Q^2}{2 \chi V_4}}
\ ,
\label{eq:gaussian-dist}
\end{align}
where $V_4=a^4\Nsite$ denotes the volume of the system under
consideration.
The following ratio of the partition functions can be written in terms
of the histogram for $\hQ$ as
\begin{align}
   {Z(\beta,\th_R+i\th_I) \over Z(\beta,0)}
&= {1 \over Z(\beta,0)}
   \sum_Q \int \!\!{\cal D}U\,
   \delta_{\hQ,Q}\,e^{-6\beta\Nsite\hsg-i(\th_R+i\th_I) \hQ}
\no\\
&= \sum_Q e^{-i(\th_R+i\th_I)Q}\,p(Q;\beta,0)
\label{eq:zratio-th}
\ .
\end{align}
Separating \eqref{eq:zratio-th} into the real and imaginary part, we
try to find the values of $\tR$ and $\tI$ where the both parts vanish
simultaneously.
Using $p(Q;\beta,0)=p(-Q;\beta,0)$, the real part is rewritten as
\begin{align}
{\rm Re}\left[ {Z(\beta,\tR+i\tI)\over Z(\beta,0)}\right]
=&\  p(0;\beta,0)
   + \sum_{Q=1}^\infty
     \cos(\tR Q)\left(e^{+\tI Q}+ e^{-\tI Q}\right)\,
     p(Q;\beta,0)
\label{eq:real-1}
\ .
\end{align}
This vanishes if $\tR$ and $\tI$ satisfy
\begin{align}
&
   \cos(\tR Q   )\, e^{- |\tI| Q   }\,p(Q  ;\beta,0)
 + \cos(\tR(Q+k))\, e^{+ |\tI|(Q+k)}\,p(Q+k;\beta,0)
 = 0
\ ,
\label{real-vanish}
\end{align}
for any integer $Q$ and any positive odd integer $k=2n+1$ with
$n=0, 1, 2, \cdots$.
This condition is rewritten as
\begin{align}
  \tI
= \pm \frac{1}{2Q+k}\ln\bigg(
- \frac{\cos(\tR Q)}{\cos(\tR (Q+k))}
  \frac{p(Q;\beta,0)}{p(Q+k;\beta,0)}
 \bigg)
\ .
 \label{eq:condition-R-1}
\end{align}
In order for the argument of the logarithm to be positive, 
\begin{align}
\tR &= (2m+1)\pi
\label{eq:tR-condition}
\ ,
\end{align}
where $m$ is any integer.
Recalling \eqref{eq:gaussian-dist}, $\tI$ is eventually found to be
\begin{align}
  \tI
= \pm \frac{2n+1}{2\chi V_4}\qquad (n=0, 1, 2, \cdots)
\ .
\label{eq:tI-condition}
\end{align}

Next, the imaginary part of \eqref{eq:zratio-th} is
\begin{align}
{\rm Im}\left[ {Z(\beta,\tR+i\tI)\over Z(\beta,0)}\right]
=& - \sum_{Q=1}^\infty
     \sin(\tR Q)\left(e^{+\tI Q}-e^{-\tI Q}\right)\,
     p(Q;\beta,0)
\label{eq:imag-1}
\ .
\end{align}
Clearly, it vanishes under \eqref{eq:tR-condition}.
Therefore, the partition function of any theories in any dimensions
vanishes at
\begin{align}
 (\tR,\ \tI)=
 \left( (2m+1)\pi,\ \frac{2n+1}{2\chi V_4} \right)
\qquad m,\ n\ : \mbox{ integer}
\label{eq:LY-zeros}
\ ,
\end{align}
as long as the theory has integer topological charge with the Gaussian
distribution, \eqref{eq:gaussian-dist}.
The facts that the zeros appear periodically in the imaginary direction
and their interval is inversely proportional to the volume tell us that
these are the Lee-Yang zeros for the first order phase transition.

Now, the question is whether the distribution \eqref{eq:gaussian-dist}
is realized in 4d SU($N$) Yang-Mills theory.
The large $N$ theory indeed shows \eqref{eq:gaussian-dist} in the large
volume limit but may not at finite volumes.
From the lattice calculation \cite{Bonati:2016tvi} studying finite
volume effects, it is expected that the distribution of $Q$  in the
large $N$ limit is, to good approximation, Gaussian even at finite
volume and thus the partition function vanishes  at \eqref{eq:LY-zeros}
to similarly good approximation.

In the large volume limit, the vacuum energy density of the large $N$
Yang-Mills theory is expected to behave with some integer
$l$~\cite{Witten:1980sp,Witten:1998uka,Bonati:2016tvi} as
\begin{align}
   \eps(\theta)
=& \min_l \frac{\chi}{2}\left(\th+2\pi l\right)^2
\label{eq:ved-largeN}
\ ,
\end{align}
which has a cusp at $\th=(2m+1)\pi$ for any integer $m$.
These singularities in the large volume limit correspond to the
(approximate) zeros at \eqref{eq:LY-zeros} at finite volume.

\section{Summary and outlook}
\label{sec:summary}

We have explored the $\th$-$T$ phase diagram of four dimensional SU(3)
Yang-Mills theory, focusing on the phase boundary, $T_c(\th)$.
Instead of measuring the Polyakov loop susceptibility, we employed the
histogram method and the constraint effective potential for the Polyakov
loop to identify the critical temperature as they provide us with other
useful information like the $\th$ dependence of robustness of the phase
transition.
Since $\th$ was introduced through the reweighting method, we could not
explore $T_c(\th)$ to $\th\sim \pi$.
The calculations succeeded to $\th \sim 0.75$ and yielded the results
for $T_c(\th)$ consistent with those
in~\cite{DElia:2012pvq,DElia:2013uaf}.

Alternatively, based on the Clapeyron-Clausius equation, one can express
$R_\th$ in \eqref{eq:def-rth} in terms of the ratio of two gaps at
$T=T_c(0)$, one being the gap of the topological susceptibility and the
other being the latent heat.
We divided configurations into the high and low temperature phases and
calculated these gaps.
For the latent heat, the precise values are available
in~\cite{Shirogane:2020muc,Borsanyi:2022xml}.
Combining these gaps, we confirmed the validity of our numerical
results.
Using those divided configurations, we also depicted the $\th$
dependence of the free energy density across the first order phase
transition.

To study possible phase transitions in the $\th$-$T$ plane from a
different point of view, we examined zeros of partition functions.
After recalling how the zeros corresponding to the deconfinement
transition appear, those associated with the spontaneous CP violation
was studied on the complex $\th$ plane, and the locations of the
Lee-Yang zeros are identified in the large $N$ limit.

There are many points to be improved in the present work.
The infinite volume limit and the continuum limit remain to be done to
bring the results obtained here to a quantitative level.
In order to approach the $\th=\pi$, it is clearly interesting to combine
the histogram method with the subvolume method~\cite{Kitano:2021jho},
which is used to calculate the vacuum energy beyond $\th=\pi$ in the
SU(2) case.
Once the whole shape of the $T(\th)$ curve on the $\th$-$T$ plane has
been determined, we would be able to gain further understandings on the
$\th$ vacuum and field theories.

\section*{Acknowledgment}

This work is supported in part by JSPS KAKENHI Grant-in-Aid for
Scientific Research (Nos.~19H00689 and 18K03662 [NY]).
This lattice code employed is based on the Bridge++ code~\cite{Ueda:2014rya}.
Numerical computation in this work was carried out in part on the
Cygnus under Multidisciplinary Cooperative Research Program (No.~17a15)
in Center for Computational Sciences, University of Tsukuba.


\appendix
\section{Applying Clapeyron-Clausius equation to the present case}
\label{app:cc}

Let $T_c(\theta)$ express the curve on the $\th$-$T$ plane, which
separates two phases by first order phase transition.
The $\th$ dependence of $T_c(\th)$ can be expressed in terms of a few
measurable quantities by applying Clapeyron-Clausius equation to this
system.
Following the argument in~\cite{Shimizu:2007}, we derive the relevant
relation below.

We define the free energy density in the high and low temperature phases
by $f_H(\th,T)$ and $f_L(\th,T)$, respectively.
Note that $f_{H}(\th,T_c(\th))=f_{L}(\th,T_c(\th))$.
We choose two points close to each other on the critical line
$T_c(\th)$, named as $(0,\ T_c)$ and $(\delta\theta,\ T_c+\delta T_c)$,
and calculate the difference of the free energy densities at these
points.
One can estimate the difference following two paths, one going through
the high temperature phase and the other through the low temperature
phase as
\begin{align}
 \mbox{path 1:}&\
f_H(\delta\th,T_c+\delta T_c) - f_H(0,T_c)
= {\partial   f_H(0,T_c)\over \partial T_c}\delta T_c +
  {1\over 2}{\partial^2 f_H(\th,T_c)\over \partial \th^2}\bigg|_{\th=0}\delta \th_c^2 +
   O(\delta^2)\ ,
\label{eq:diff-H}
\\
 \mbox{path 2:}&\
f_L(\delta \th,T_c+\delta T_c) - f_L(0,T_c)
= {\partial   f_L(0,T_c)\over \partial T_c}\delta T_c +
  {1\over 2}{\partial^2 f_L(\th,T_c)\over \partial \th^2}\bigg|_{\th=0}\delta \th^2 +
   O(\delta^2)\ ,
\label{eq:diff-L}
\end{align}
where $O(\delta^2)$ representatively expresses $O(\delta T_c^2)$,
$O(\delta\th^4)$ or $O(\delta T_c\delta\th^2)$.
Note that the symmetry of $f_{H,L}(-\th,T)=f_{H,L}(\th,T)$ is used
above.

Recalling $f=\epsilon-Ts$ with $\epsilon$ internal energy density and
$s$ entropy density and $\partial^2 f/\partial\th^2|_{\th=0}=\chi$
and using the fact that the difference \eqref{eq:diff-H} and
\eqref{eq:diff-L} are equal, the following holds up to $O(\delta^2)$,
\begin{align}
  - s_H(0,T_c)\delta T_c + {1\over 2}\chi_H \delta\th^2
\approx
  - s_L(0,T_c)\delta T_c + {1\over 2}\chi_L \delta\th^2
\ .
\end{align}
Using the latent heat,
$\Delta\epsilon=(s_H - s_L ) T_c=\epsilon_H-\epsilon_L$, and noticing
that $T_c(\delta\th)/T_c(0)=1+\delta T_c/T_c$, one ends up with
\begin{align}
 {T_c(\th)\over T_c(0)}
\approx
 1 - {\chi_L(T_c)- \chi_H(T_c) \over 2 \Delta\epsilon} \th^2
\ .
\label{eq:cc-relation}
\end{align}
This relation is derived also in \cite{DElia:2013uaf} in a similar way.

\end{document}